\documentclass[preprintnumbers,secnumarabic,amsmath,amssymb,nofootinbib,floatfix,superscriptaddress]{revtex4}
\usepackage{varioref,exscale,latexsym,amsmath,amssymb}
\usepackage{graphicx,epstopdf}
\usepackage{comment}  
\usepackage{color}

\usepackage{epsfig}
\usepackage{graphicx}
\usepackage{dcolumn}
\usepackage{bm}
\usepackage{simplewick}
\usepackage{hyperref} 

\begin{document}

\title{{\bf A rapid holographic phase transition with brane-localized curvature}}

\medskip\
\author{Barry Dillon}
\email[Email: ]{barry.dillon@plymouth.ac.uk}
\affiliation{Centre for Mathematical Sciences,
Plymouth University\\
Plymouth, PL4 8AA, United Kingdom}
\author{Basem Kamal El-Menoufi}
\email[Email: ]{b.elmenoufi@sussex.ac.uk}
\author{Stephan J.~Huber}
\email[Email: ]{s.huber@sussex.ac.uk}
\author{Jonathan P.~Manuel}
\email[Email: ]{j.manuel@sussex.ac.uk}
\affiliation{Department of Physics and Astronomy,
University of Sussex\\
Falmer, Brighton, BN1 9QH, United Kingdom}

\begin{abstract}
We study the finite-temperature properties of the Randall-Sundrum model in the presence of brane-localized curvature. At high temperature, as dictated by AdS/CFT, the theory is in a confined phase dual to the planar AdS black hole. When the radion is stabilized, {\em \' a la} Goldberger-Wise, a {\em holographic} first-order phase transition proceeds. The brane-localized curvature contributes to the radion kinetic energy, which allows us to substantially decrease the critical bubble energy. Contrary to previous results, the phase transition completes at much larger values of N, the number of degrees of freedom in the CFT. Moreover, the value of the bulk scalar on the TeV-brane is allowed to become large, consistent with back-reaction constraints. Assisted by this fact, we find that for a wide region in the parameter space tunneling happens rather quickly, i.e. $T_n/T_c \sim \mathcal{O}(0.1-1)$. At zero temperature, the most important signature of brane-localized curvature is the reduction of spin-2 Kaluza-Klein graviton masses and a heavier radion.
\end{abstract}
\maketitle
\flushbottom

\section{Introduction}
	
	The possibility that physical reality might consist of extra spatial dimensions has intrigued theoretical physicists since at least the work of Kaluza and Klein. The modern excitement about extra dimensions re-surfaced when it was realized they offer a simple solution to the hierarchy problem \cite{Ark1998}. Most prominent is the proposal by Randall and Sundrum \cite{Ran1999} in which the bulk of spacetime is a slice of AdS$_5$ bordered by two branes in which, unlike gravity, matter is confined to propagate only on the TeV brane. The hierarchy between the weak and Planck scales dynamically emerges due to the effect of gravitational redshift: mass scales on the TeV brane are exponentially redshifted by the warp factor.  	
	
	The cosmology of extra-dimensional models came under scrutiny immediately after their proposal. In the absence of brane cosmological constants, the extra dimension leads to unconventional cosmology on the brane, i.e. $H \propto \rho$~\cite{Bin1999}. Nevertheless, in the Randall-Sundrum (RS) model standard cosmology could be recovered at low energy but at the cost of requiring finely-tuned energy densities on the two branes \cite{Cli1999} and negative energy matter content in our Universe. This fine tuning is lifted if the extra dimension is dynamically stabilized \cite{Csa1999}, yet, deviation from standard cosmology ultimately takes place at early times above $ \sim 1\, \text{TeV}$ \cite{Les2000}. The above problems evidently prompt questions about the high temperature dynamics of the Randall-Sundrum I (RSI) model.  
	
	  Creminelli et al. sought to address these issues by employing the AdS/CFT correspondence \cite{Cre2001}. The holographic description of the RSI model has been discussed extensively in the literature \cite{Ark2000,Rat2000}. Most importantly, the dual strongly coupled conformal field theory (CFT) is {\em confining} where the spin-2 bound states are conjectured to be dual to the Kaluza-Klein (KK) spin-2 spectrum of the 5D theory. At high temperature, as familiar from QCD, one should expect the theory to undergo a phase transition and enter a {\em de-confined phase}. In fact, Witten \cite{Wit19982} considered a similar situation using the AdS/CFT dictionary to study the phase transition of a large-$N$ gauge theory defined on $\mathbb{S}^3$. On the gravity side, the high temperature phase is the AdS-Schwarzschild black hole while the low temperature phase is thermal AdS. This is the well-known Hawking-Page phase transition \cite{Haw1982}.
	  
	  The physics in our situation is now transparent: at high temperatures, the RSI is unstable and the theory must undergo a phase transition. Similar to the Hawking-Page phase transition, the high temperature phase is described by the planar AdS black hole \cite{Hor1998}. However, unlike Hawking-Page, the planar AdS black hole is stable at any finite temperature {\em unless} the extra dimension in the RSI solution is dynamically stabilized \cite{Cre2001}. The metric fluctuations contain a scalar degree of freedom, i.e. the radion, that controls the physical size of the extra dimension. A stabilization mechanism, e.g. Goldberger-Wise \cite{Gol1999}, generates a potential for the radion that renders the phase transition possible as we shall see below.
	  
	  Several works have been devoted to study the dynamics of the holographic phase transition \cite{Cre2001,Ran2006,Nar2007,Kon2010} and, albeit introducing various twists, two general properties of the phase transition (PT) were found to hold true regardless of the details. First, the PT is strongly first order with substantial amounts of supercooling. Second, the system gets stuck in the false vacuum if the ratio of the 5D Planck mass to the bulk cosmological constant is larger than $\mathcal{O}(1)$, taking the best case scenario \cite{Kon2010}. Although it might be desirable for certain applications to have large amounts of supercooling, the second property threatens the framework consistency, in particular, the holographic interpretation of the transition. Effective field theory lore tells us that the Einstein-Hilbert action dominates the dynamics on the gravity side only if the ratio of the 5D Planck mass to the cosmological constant is large. Precisely, AdS/CFT asserts the correspondence \cite{Mal1997}
	  \begin{align}\label{Ndef}
	  N^2 = 4 \pi^2 (M_\star l)^3 + 1 \ \ ,
	  \end{align} 
	  where $M_\star$ is the 5D Planck mass, $l$ is the AdS radius and $N$ is the number of CFT degrees of freedom. Ideally, for the classical gravity description to be meaningful, one must have $(M_\star l) \gg 1$ or large $N$. At large $N$, the 4D field theory is strongly coupled which is essential for understanding the phase transition.  
	  
	 In this paper, we consider a minimal extension to the RSI set-up, namely a TeV-brane localized curvature \cite{Dav2003}
	 \begin{align}\label{braneterm}
 	S^{\text{G}}_{\text{IR}} = \frac{M_{\text{IR}}^2}{2} \int d^4x \sqrt{g_L}\, R(g_L)
 	\end{align} 
	 where $M_{\text{IR}}$ is a mass scale comprising the new parameter in the model, $g_L$ is the induced metric on the TeV brane and $R$ is the Ricci scalar built from the induced metric. Naturally, one should expect $M_{\text{IR}}$ to be of the same order of magnitude as other mass scales in the model, i.e. the 5D Planck mass and the bulk cosmological constant. 
  We are concerned with the finite temperature aspects of the model, in particular the dynamics of the holographic phase transition. We shall see that the kinetic energy of the radion receives a non-trivial modification due to Eq.~\eqref{braneterm} and allows for a large-$N$ phase transition to proceed over a wide range of parameters appearing in the radion potential. Moreover, the model accommodates a wide range of nucleation temperatures: A large amount of supercooling is achieved if the mass-squared of the bulk scalar field is small compared to the AdS curvature. 
	 	 	 
	 The plan of the paper is as follows. In section \ref{sect2}, we define the radion and derive its kinetic Lagrangian employing the orbifold formulation of the RSI model. We then move to construct the radion potential induced via the Goldberger-Wise mechanism. Section \ref{sect3} is devoted to determining the free energies and discussion of the modeling of the phase transition and its holographic interpretation. Section \ref{sect4} offers a detailed study of the dependence of the tunneling rate on the various model parameters paying special attention to the role brane-localized curvature plays in the phase transition. The phenomenological consequences of Eq.~\eqref{braneterm} and possible collider signatures of the radion are the subject of section \ref{sect5}. Our conclusions are given in section \ref{sect6}.

\section{Effect of brane-localized curvature on the radion}\label{sect2}

In this section we derive the effective 4D Lagrangian for the radion including the brane-localized curvature. The radion is a scalar degree of freedom that controls the physical size of the extra dimension. The brane-localized curvature contributes to the kinetic energy of the radion, thereby introducing a non-trivial modification to its effective Lagrangian. In the absence of a stabilization mechanism the radion is strictly massless. We employ a bulk scalar field, {\em \' a la} Goldberger and Wise~\cite{Gol1999}, to generate a potential for the radion. We review the computation of the potential starting from the Charmousis-Gregory-Rubakov (CGR) wave-function \cite{Cha1999}.

	\subsection{Radion kinetic term}\label{radionKE}
	
	It is important to derive the radion kinetic term in detail because it is instructive to point out the difference in the result obtained from the orbifold picture of the RSI model to that obtained in the so-called interval approach~\cite{Car2005,Geo2011}. In the orbifold picture, the action of the theory is composed of bulk and boundary parts as
	\begin{align}\label{gravitybulk}
	S^{\text{G}}_{\text{bulk}} = \frac{M_\star^3}{2} \int d^4x \int_{-L}^{L} dy\, \sqrt{G} \left( R + 12\, k^2 \right), \quad 	k^2 		\propto \Lambda
	\end{align}
and
	\begin{align}\label{gravitybound}
	S^{\text{G}}_{\text{B$_{i}$}} = - \sigma_i \int d^4x  \int_{-L}^{L} dy\, \sqrt{g_i} \,\delta(y-y_i), \quad y_i =0, L
	\end{align}
	where the Planck (TeV) brane is located at the orbifold fixed point $y=0 \,(L)$ and $g_i$ denotes the induced 4D metric on the branes. The extra dimension is compactified on an $\mathbb{S}^1/Z_2$ orbifold and the theory admits a static solution
	\begin{align}\label{staticRS}
	ds^2 = e^{-2 k |y|} \eta_{\mu\nu} dx^\mu dx^\nu - dy^2
	\end{align}
	 provided the brane tensions are fine-tuned to obtain flat brane geometries
	\begin{align}
	\sigma_0 = - \sigma_L = 6k M_\star^3 \ \ .
	\end{align}
	The background solution remains intact even with the introduction of the brane-localized curvature 
	\begin{align}\label{branecurv}
 	S^{\text{G}}_{\text{IR}} = \frac{M_{\text{IR}}^2}{2} \int d^4x \sqrt{g_L}\, R(g_L)  \ \ ,
 	\end{align} 
	as one could easily see by noticing that the induced metric on the TeV brane is flat. The 4D Planck mass has a dependence on the new mass scale, $M_\text{IR}$, as
	\begin{equation}\label{4dplanck}
	M^2_{\text{Pl}}=\frac{M_\star^3}{k}\left( 1-e^{-2kL}+e^{-2kL}\, \frac{M^2_{\text{IR}}k}{M^3_\star} \right)\ ,
	\end{equation}
	where, with $e^{-2kL}\ll 1$, the dependence is indeed very mild.
	
It is useful to pause and comment on the holographic description of brane-localized curvature. The RS models are thought to be dual to strongly coupled 4D theories that are approximately conformal between $M_\star$ and the Kaluza-Klein scale.
The presence of UV and IR branes explicitly breaks conformal symmetry, and matter fields on the UV (IR) brane are seen as dual to fundamental (composite) fields in the dual theory. Hence, the massless graviton (UV localized) is thought to be mostly fundamental, while the Kaluza-Klein gravitons (IR localized) are thought to be mostly composite. In the absence of the IR brane ($L \rightarrow \infty$) the dual theory would be conformal at all energies below $M_\star$, and the effective Planck mass would be $M_{\text{Pl}}^2=M_\star^3/k$.
In the dual theory the bulk Einstein-Hilbert action arises dynamically via loops of CFT fields cut off at $M_\star$, while the spontaneous breaking of conformal invariance, due to the presence of the IR brane, generates a $-e^{-2kL}M_\star^2/k$ correction to $M^2_{\text{Pl}}$.
Now, the presence of an IR brane kinetic term also corrects the effective Planck mass, Eq.~(\ref{4dplanck}), thus in terms of the dual theory it should be thought of as arising due to modified dynamics associated with the breakdown of conformal invariance in the IR.
For a more thorough discussion on the CFT interpretation we refer the reader to \cite{Dillon:2016bsb}.

	Formally, the radion is the scalar zero-mode in the metric perturbations. Physically, radion fluctuations control the size of the extra dimension. It was shown in Ref.~\cite{Cha1999} that an appropriate ansatz for scalar perturbations is
	\begin{align}\label{scalarpert}
	ds^2 = e^{-2ky-2F} \eta_{\mu\nu} dx^\mu dx^\nu - (1 + 2F)^2 dy^2, \quad F \equiv F(x^\mu,y)
	\end{align}
	 where $F$ is related to the physical radion field as we show below. In the absence of Eq.~\eqref{branecurv}, one could determine the radion wave-function by either solving the linearized Einstein equations or simply working on the action level. For example, one could expand the actions in Eqs.~\eqref{gravitybulk} and~\eqref{gravitybound} to quadratic order in $F(x^\mu,y) \equiv f(y) R(x)$. The wave-function is then determined by simply demanding the quadratic fluctuations to be massless, which yields the following differential equation for the wave-function in the bulk
	 \begin{align}
	 400 k^2 f^2 + 48 f \ddot{f} + 36 \dot{f}^2 - 368 k f \dot{f} = 0
	 \end{align}
	 with the following unique solution respecting the orbifold boundary conditions\footnote{The brane terms resulting from Eq.~\eqref{gravitybound} force the boundary conditions, automatically satisfied by Eq.~\eqref{radionwfunc}.}:
	 \begin{align}\label{radionwfunc}
	 f(y) = e^{2 k |y|}  \ \ .
	 \end{align}
	 
	 The same solution is recovered from the equations of motion; see, e.g., Ref.~\cite{Cha1999}. The situation is drastically different in the presence of brane curvature: Eq.~\eqref{radionwfunc} does {\em not} solve the linearized equations of motion. Nevertheless, this poses no concern. First, the theory is ultimately defined via a path integral that only employs the action as the fundamental object. Second and more importantly, the wave-function in Eq.~\eqref{radionwfunc} is physically motivated as it yields a degree of freedom whose value measures the size of the extra dimension.
	 
	To derive the kinetic term of the radion, there is a subtle issue that we wish to point out. The available derivation in the literature starts by first expanding the action to quadratic order in fluctuations, i.e. $F(y,x^\mu)$, and proceeds using the CGR wave-function to integrate along the extra dimension thereby obtaining the usual kinetic term. In fact, this procedure is inaccurate for the purpose of studying the phase transition and is adequate only when discussing radion phenomenology at zero temperature. In a nutshell, the procedure of expanding the action to second order in $F$ leads to a non-standard kinetic term of the form $(\partial \mu)^2/\mu^2$. As we shall see below, this problem is artificial and an exact computation yields a proper kinetic term for the radion.
			
	Now we plug Eq.~\eqref{scalarpert} into \eqref{gravitybulk} without any expansion. A straightforward computation yields
	\begin{align}\label{gravitybulkF}
	 S^{\text{G}}_{\text{bulk}} = \frac{M_\star^3}{2} \int d^4x \int_{-L}^{L} dy \left[6 e^{-2 k |y|} e^{-2F} (1-2F) (\partial_\alpha F)(\partial^\alpha F) - 8 k^2 (1-2F)\, e^{- 4 k |y|} e^{-4F} \right]
	 \end{align} 
	where we have integrated by parts and utilized Eq.~\eqref{radionwfunc} to considerably simplify the term inside brackets. Because of orbifold boundary conditions, there are terms in the action proportional to delta functions but they all cancel identically with the brane actions in Eq.~\eqref{gravitybound}. Lastly, we need to carry out the integral over $y$ which prompts us to recall the definition of the radion field. Precisely,
	\begin{align}
 	\mu(x) \equiv k e^{-kd(x)} 
 	\end{align}
	where $d(x)$ is the proper length of the extra dimension
	\begin{align}
	k d(x) = k \int_0^L dy (1+2F) = k L + R(x) (e^{2kL} - 1) \ \ .
	\end{align}
	
	Now the integral in Eq.~\eqref{gravitybulkF} is performed via a change of variables
	\begin{align}
	k\bar{y} = ky - e^{2ky} R(x) + R(x)
	\end{align}
	and one finds 	
	\begin{align}
 	\mathcal{L} =  \frac{6 M_\star^3}{k^3} \,  \frac12 (\partial_\alpha \mu)\, (\partial^\alpha \mu)
 	\end{align}
	where we used that $e^{2kL} \gg 1$ and ignored a self-interaction term $\mathcal{O}(\mu^4)$. Adding Eq.~\eqref{branecurv} now leads to our main result
	\begin{align}\label{znorm}
	\mathcal{L} = \mathcal{Z}^2 \,  \frac12 (\partial_\alpha \mu)\, (\partial^\alpha \mu)
	\end{align}
	where 
	\begin{align}
	\mathcal{Z} \equiv  \sqrt{\frac{6 M_\star^3}{k^3}  (1-\theta_{\text{IR}})}, \quad \theta_{\text{IR}} \equiv \frac{M^2_{\text{IR}} k}{M_\star^3} \ \ .
	\end{align}
	
	Notice here that the sole effect of Eq.~\eqref{branecurv} is to modify the kinetic energy of the physical radion. Indeed, we demand that $\theta_{\text{IR}} < 1$ to insure the radion is not a ghost. Our result differs from that obtained by employing the interval approach~\cite{Geo2011}, which hinges on solving the linearized equations of motion in the presence of Eq.~\eqref{branecurv}. In addition, notice the interval approach parametrizes the scalar perturbations differently than Eq.~\eqref{scalarpert}. We stick to the CGR wave-function, and the corresponding kinetic term, since it allows for a transparent relation between $F(x,y)$, the radion and the size of the extra dimension.

	\subsection{Stabilization and the radion potential}
	
	We shall see in the next section that the high temperature phase of the theory, holographically dual to the planar AdS black hole, is stable at any temperature and no phase transition can occur. This completely changes once the extra dimension in the Randall-Sundrum set-up is stabilized. Indeed even in the zero-temperature realization of the model, stabilization is required to invoke a dynamical mechanism generating the large hierarchy between the weak and Planck scales in a natural way. 
	
	The mechanism of Goldberger and Wise is minimal and serves our purposes. It relies on adding a bulk scalar field that develops a non-trivial profile along the extra dimension. Using this profile in the Goldberger-Wise action, a potential for the radion field is generated with a global minimum thus dynamically fixing the size of the extra dimension.
	
	Here, we derive the radion potential using the correct metric ansatz for the scalar perturbation. In the bulk, the scalar field equation is
	\begin{align}
	(\Box + m^2) \varphi(y) = 0
	\end{align}
	which is evaluated using the metric in Eq.~\eqref{scalarpert} and leads to
	\begin{align}
	(1+2F) \varphi^{\prime\prime} - 4 (1+2F)(k+F^\prime) \varphi^\prime - 2 F^\prime \varphi^\prime = (1+2F)^3 m^2 		\varphi \ \ , 
	\end{align}
	where a prime denotes derivative with respect to $y$. Remarkably, the bulk equation could be solved exactly with the coordinate transformation
	\begin{align}
	k\bar{y} = k y + e^{2ky} R(x) - R(x)
	\end{align}
which turns the equation simply to 
	\begin{align}
	\ddot{\varphi} - 4 k \dot{\varphi} = m^2 \varphi, \quad \dot\varphi \equiv \partial_{\bar{y}} \varphi \ \ .
	\end{align}

We see that the coordinate transformation has changed the independent variable to be the proper distance along the extra dimension, i.e. $\bar{y}$. Now, the solution is rather simple,
\begin{align}
 \varphi(\bar{y}) = e^{2k\bar{y}} \left[A e^{\nu k \bar{y}} + B e^{-\nu k \bar{y}} \right], \quad \nu \equiv \sqrt{4 + \frac{m^2}{k^2}}
\end{align}
where $A$ and $B$ are integration constants to be fixed by boundary conditions. The exact forms of the boundary actions do not concern us as they merely enforce the boundary conditions. We can easily switch back to the coordinate $y$ to determine the constants
\begin{align}
\varphi(y=0) = \Phi_P\ , \quad \varphi(y=L) = \Phi_T \ \ .
\end{align}

The potential of the radion is determined by plugging the solution back into the action and integrating over the extra dimension. Precisely, we have
\begin{align}
V(\mu) = \int_0^L \,dy\  e^{-4ky - 4F} (1+2F) \left[(1+2F)^{-2} \varphi^{\prime 2} + m^2 \varphi^2 \right]\ ,
\end{align}
	  which is evaluated most simply by switching to $\bar{y}$ to yield
	  \begin{align}
	V(\mu) = k(\nu+2) A^2 \big(e^{2\nu k d(x)} - 1\big) + k(\nu-2) B^2 \big(1-e^{-2k\nu d(x)}\big)\ ,
	\end{align}
where
	\begin{align}
	A + B = \Phi_P\ , \quad A e^{\nu k d(x)} + B e^{-\nu k d(x)}  = e^{-2 k d(x)} \Phi_T \ \ .
	\end{align}
	
	So far, no approximations have been made. We can follow Goldberger and Wise to solve for the constants in the limit of large extra dimension, i.e. $k d(x) \gg 1$, which indeed is a valid limit given that we want to solve the hierarchy problem. Another convenient limit is that of small $\epsilon \equiv m^2/4k^2$ that enables the extraction of analytic results. In terms of the physical radion, we finally have
	\begin{align}\label{radionpot}
	V(\mu)  = \epsilon k^4 \Phi_P^2 + \mu^4 \left[(4+2\epsilon) \left(\Phi_T-\Phi_P \frac{\mu^\epsilon}{k^\epsilon}\right)^2 - 		\epsilon \Phi_T^2 \right]
	\end{align}
where the brane vacuum expectation values (vevs) are made dimensionless via
	\begin{align}
	\Phi_P \to k^{3/2} \Phi_P, \quad \Phi_T \to k^{3/2} \Phi_T \ \ .
	\end{align}

We shall thoroughly discuss the constraints on the brane vevs in section \ref{sect4}. This completes our derivation of the radion effective Lagrangian.

\section{Thermal phase transition: holographic description}\label{sect3}

It is mandatory to study the finite temperature behavior of the RS model if it were to describe the early Universe. This endeavor was initiated in Ref.~\cite{Cre2001} concluding that finite-temperature effects tend to destabilize the extra dimension. At low enough temperature (below a TeV) and in the absence of a stabilization mechanism, a temperature-dependent potential for the radion is induced through its coupling to the heavy KK gravitons. When the latter are integrated out and a thermal average of the effective action is performed, a correction of the form $(-T^8/\mu^4)$ appears~\cite{Cre2001}. Clearly, the latter correction renders the RS set-up unstable by pushing the branes away from each other.  

The authors of Ref.~\cite{Cre2001} turned to the AdS/CFT correspondence to understand the high temperature phase of the model. The RSI is conjectured to be dual to a strongly coupled CFT~\cite{Ark2000,Rat2000}. The introduction of branes is then seen to break conformal symmetry. The Planck brane explicitly breaks the symmetry by introducing a UV cut-off while coupling the CFT to 4D gravity. The TeV brane, on the contrary, signals the spontaneous breakdown of the symmetry in the IR where the scale of symmetry breaking is set precisely by the size of the extra dimension. Matter fields on the TeV brane, as well as KK states, are dual to bound states in the 4D theory.

The finite temperature physics of the RSI model becomes very transparent if one recalls the exact AdS/CFT dictionary \cite{Wit1998}
\begin{align}
Z_{\text{CFT}}[g] = Z_{\text{gravity}}[G]
\end{align}
where the rhs denotes the thermal partition function of a quantum theory of gravity formulated on AdS space with metric $G$ that induces a boundary metric $g$. In the saddle-point approximation to the partition function, one finds all solutions to the Euclidean Einstein equations representing thermal equilibrium. The partition function is then a sum over these disconnected saddles, ideally taking loop fluctuations into account. In our case, there are two solutions that represent states of thermal equilibrium; the RSI model with two branes and the planar AdS black hole. This observation, first made in Ref.~\cite{Cre2001}, is key to unveiling the high temperature characteristics of the model, and the physics turns out to be quite similar to the Hawking-Page phase transition.  

Let us recall the AdS planar black hole \cite{Hor1998}. The metric reads
\begin{align}
ds^2 = \frac{r^2}{l^2} \left(1-\frac{r^{D-2}_0}{r^{D-2}}\right) dt^2 - \frac{l^2}{r^2} \left(1-\frac{r^{D-2}_0}{r^{D-2}}\right)^{-1} dr^2 - \frac{r^2}{l^2} \sum_{i=1}^3 dx_i^2
\end{align}
where the transverse dimensions, i.e. $x_i$, could be made compact by restricting the range of coordinates. There exists a single horizon at $r=r_0$ with $\mathbb{R}^3$ topology. The parameter $r_0$ is related to the ADM mass per unit transverse volume
\begin{align}
\frac{\mathcal{M}}{V} = \frac{3M_\star^3 r_0^4}{2l^5} \ \ .
\end{align}

Armed with the two solutions, we need to compare their respective free energies to decide on the relative stability of each phase. The Euclidean approach to black hole thermodynamics commences with introducing the Euclidean section via an analytic continuation $t \to -i\tau$. The periodicity of imaginary time is fixed so as to avoid a conical singularity~\cite{Gib1976}, i.e.
\begin{align}
\beta = \frac{\pi l^2}{r_0} \ \ .
\end{align}

The contribution of the saddle-point to the free energy is given by the Euclidean action, i.e. $\beta F(\beta) = \mathcal{S}_E$. The free energy of the black hole diverges as $r \to \infty$, so we subtract off the pure AdS contribution 
\begin{align}\label{freepBH}
F_{\text{pBH}} - F_{\text{AdS}}  = -\frac{\pi^4}{2} (M_\star l)^3 T^4 \ \ .
\end{align}
Notice that the temperature of AdS is not fixed by any requirement. The relation between the black hole and AdS temperatures, as devised by Hawking and Page~\cite{Haw1982}, is found by equating the proper length of the thermal circle at spatial infinity. The conclusion from Eq.~\eqref{freepBH} is very elegant: unlike the AdS-Schwarzschild black hole, the planar black hole is always stable and never decays to thermal AdS. 
	
	The free energy of the RSI solution is more subtle. Simply put, the computation yields a finite result while thermal AdS yields a divergent contribution. The situation is easily remedied by  simply sending the Planck brane to infinity to coincide with the AdS boundary. To see this, we introduce a new coordinate, $ky = -\ln (kr)$, which turns Eq.~\eqref{staticRS} into
	\begin{align}
	ds^2 = \frac{r^2}{l^2} dt^2 - \frac{l^2}{r^2} dr^2 - \frac{r^2}{l^2} \sum_{i=1}^3 dx_i^2
	\end{align}
 	which is the same as AdS except that $r \in [r_{\text{TeV}},\infty)$. Upon including the Hawking-Gibbons-York boundary term for the TeV brane, one simply finds
\begin{align}\label{freepRS}
F_{\text{RS}} - F_{\text{AdS}}  = 0
\end{align}
	which means again that no phase transition could proceed and the theory stays in the de-confined phase at any temperature. Indeed, there is a caveat to this result: the situation drastically changes once stabilization is taken into account upon adding the bulk scalar. The simplest way to see this is by inspecting the induced radion potential in Eq.~\eqref{radionpot} which contributes to the free energy of the RSI phase inciting a phase transition at a critical temperature
	\begin{align}\label{Tcrit}
	T^4_{\text{c}} &= \frac{2V(\mu_-)}{\pi^4 (M_{\star} l)^3} 
	\end{align} 
where $\mu_-$ denotes the radion vev.

In the five-dimensional description used here, the phase transition is characterized by a jump in the radion field, which therefore takes the role of the order parameter. As we explained, the five-dimensional phase transition corresponds in the dual theory to a (de)-confinement transition of the strongly coupled theory. At high temperatures, the system has a nearly conformal symmetry, which spontaneously breaks when the strong sector confines. This symmetry breakdown will generate a composite pseudo-Goldstone boson, i.e. the dilaton. The vev of the dilaton field sets the confinement scale, and so can be interpreted as the order parameter on the four-dimensional side \cite{Bru2018}.


\section{Dynamics of the phase transition at large N}\label{sect4}
We saw in the last section that the stabilization-induced radion potential triggers a first order phase transition to proceed from the high temperature black hole phase to the low temperature RSI phase. In this section, we study the dynamics of the phase transition. First, we illuminate how the various parameters of the model affect the features of the phase transition. Second, we discuss the effect of turning on $\xi_{\text{IR}} \equiv 1-\theta_{\text{IR}}$.

	\subsection{Qualitative discussion}

\begin{figure}[tbp]
	\center
  	\includegraphics[scale=0.7]{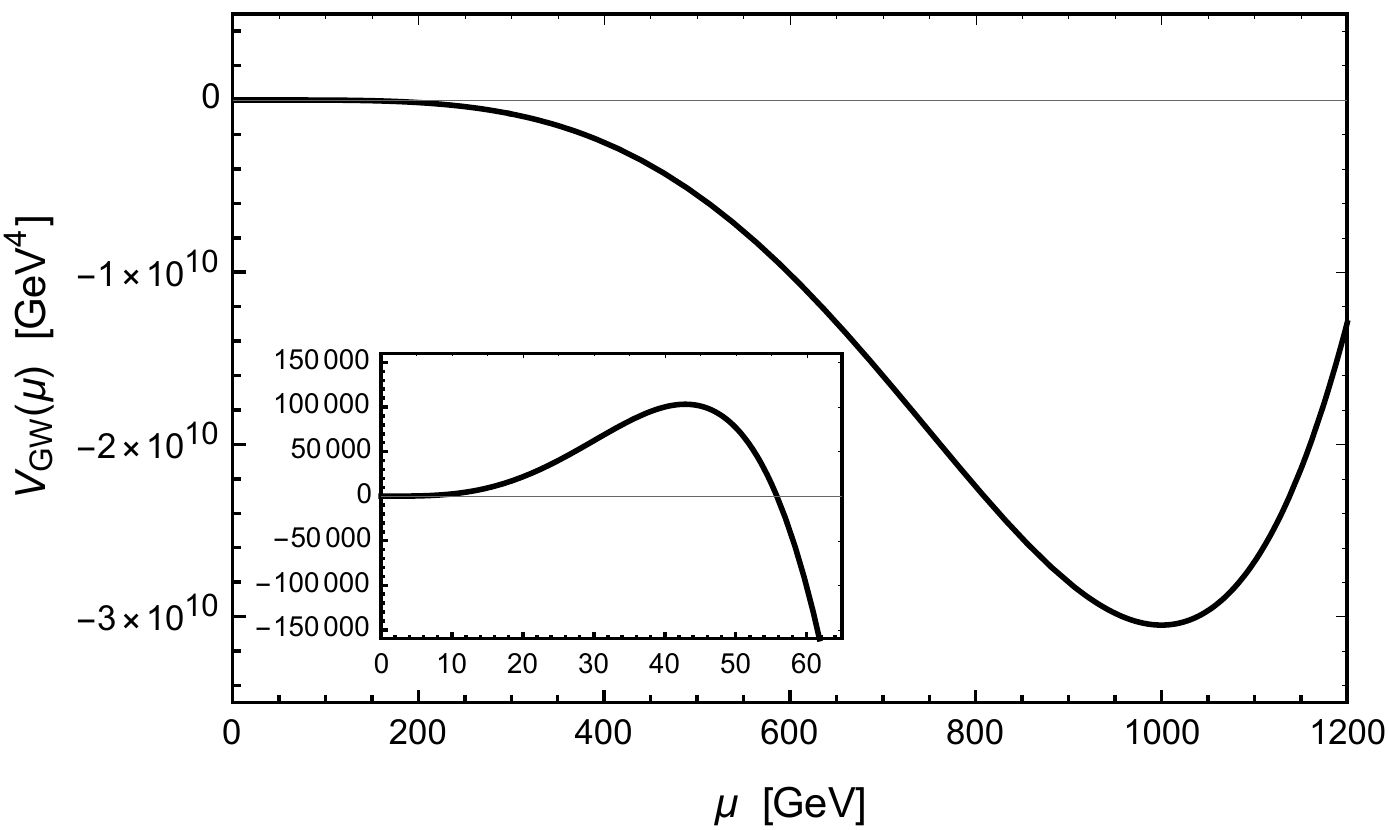}
  	\caption{A plot of the radion potential, described by Eq.~\eqref{finLag}, for $\Phi_T=1,\ \epsilon=0.1$ and $(M_\star l) = 0.55$. The inset shows the size and location of the maximum. \label{potentialplot}}
	\end{figure}

	It is instructive at this stage to discuss in detail the effect various parameters have on the strength of the phase transition. To this end, we start by rewriting the radion Lagrangian 
	\begin{align}\label{finLag}
	\mathcal{L} = \mathcal{Z}^2\, \frac12 (\partial \mu)^2  -   \Phi_T^2 \left[(4+2\epsilon) \left(1 - (\Phi_P/\Phi_T) \frac{\mu^\epsilon}{k^\epsilon}\right)^2 - \epsilon \right] \mu^4 \ \ ,
	\end{align}
	where we explicitly uncover that $\Phi_T$, unlike $\Phi_P$, plays a dominant role in the dynamics. The generic features of the potential are displayed in Figure~\ref{potentialplot}. Most importantly, for any positive $\epsilon$ the potential possesses a global maximum and a minimum. The values of $\Phi_T$ and $\epsilon$ predominantly control both the position of, as well as the value of the potential at the extrema.
		
	 As it stands, this Lagrangian is still not adequate to discuss the tunneling process. We need to subtract off the energy of the false vacuum, i.e. Eq.~\eqref{freepBH}, which serves as a temperature-dependent offset when computing the bubble action. This leaves the potential
	\begin{equation}\label{radionpotoffset}
	V(\mu,T)  = \Phi_T^2 \left[(4+2\epsilon) \left(1-(\Phi_P/\Phi_T) \frac{\mu^\epsilon}{k^\epsilon}\right)^2 - \epsilon \right] \mu^4 + \frac{\pi^4}{2}(M_\star l)^3T^4\ .
	\end{equation}
		 
	 In conventional tunneling problems the potential describes the whole scenario and one uses a bounce solution to find a bubble profile that ends in the false vacuum with zero kinetic energy. The case at hand is drastically different since, as we explained, we do not consider the contribution of the black hole to the tunneling rate and our potential therefore only describes the dynamics of the radion. To handle this problem, we follow the approach used by Ref.~\cite{Kon2010} which tacitly takes care of this issue. As in the usual bounce solution approach the radion is released somewhere between the extrema of the potential and is then required to tunnel to $\mu=0$. However, given that $\mu=0$ is not the false vacuum we demand the radion field reaches this point with enough kinetic energy to jump over and mount on top of the black hole free energy, the correct false vacuum. Precisely, the bubble profile now obeys the boundary condition \cite{Kon2010}
	\begin{align}\label{BC}
	\xi_{\text{IR}}\, \frac12 (\partial_\alpha \mu)(\partial^\alpha \mu)\vert_{\mu=0}  = \frac{\pi^4}{12} T^4 \ \ ,
	\end{align}
	which consequently introduces a non-trivial temperature dependence in the tunneling rate in addition to the temperature dependent offset now in the potential. Once the parameters $(M_\star l),\, \Phi_T$ and $\epsilon$ are fixed, Eq.~\eqref{BC} associates each release point $\mu_r$ to a temperature value. We believe the treatment outlined in Ref.~\cite{Kon2010} is the most suitable in our situation given our inability to include the black hole phase in computing the tunneling rate. 
	
	Radion field configurations are found by using the boundary condition, Eq.~\eqref{BC}, in conjunction with solutions of the canonically normalized $O(3)$ equation of motion
	\begin{equation}\label{O3_eom}
	\partial_r^2 \mu + \frac{2}{r}\cdot\partial_r \mu = \frac{1}{\mathcal{Z}^2}\cdot\frac{\partial V(\mu,T)}{\partial\mu}
	\end{equation}
	where $\mathcal{Z}$ and $V(\mu,T)$ are given by Eqs.~\eqref{znorm} and \eqref{radionpotoffset} respectively, although note that the temperature dependent offset plays no role here. Eq.~\eqref{O3_eom} is solved for a range of release points $\mu_r$ which in turn translate to a range of temperatures through Eq.~\eqref{BC}. The solutions and corresponding temperatures are then used to find the $O(3)$ bounce action from
	\begin{equation}
	\frac{S_3}{T} = \frac{4\pi}{T}\cdot\mathcal{Z}^2\cdot\int dr\ r^2 \left[ \frac{1}{2}(\partial_r\mu)^2 + \frac{1}{\mathcal{Z}^2}\cdot V(\mu,T)\right]\ .
	\end{equation}
	
	The tunneling rate per unit volume reads
	\begin{align}
	\Gamma = A(T)\, e^{-S_3/T},
	\end{align}
	where by dimensional analysis we shall estimate $A(T) \sim (\mu_-)^4 $. The requirement of nucleating one critical bubble per unit Hubble volume then translates into
		\begin{align}\label{S3T}
		 \frac{S_3}{T} \leq 4 \log\left(\frac{M_P}{T_c} \right) + 4 \log\left(\frac{\mu_-}{T_c} \right),
		 \end{align}
	where the cosmology is radiation-dominated. Hence, if we take $T_c$ to be of the same order as $\mu_-$, we obtain $S_3/T \lesssim 140$. This is the value we use throughout our study although a lower $T_c$, attainable in most of our parameter space, considerably relaxes the nucleation condition.
	
	Including the false vacuum offset, it is illustrative to split the bounce action into two contributions
	\begin{align}\label{s3overTdecomp}
	\frac{S_3}{T} = \frac{E}{T} + \frac{2 \pi^5}{3} (M_\star l)^3 \left(r_c T\right)^3 \ \ ,
	\end{align}
	 where $E$ and $r_c$ are, respectively, the energy and radius of the critical bubble. In such a form it becomes apparent that the factors influencing the bounce action are $E$, $T$, and $r_c$. Each of these factors has a non-trivial dependence on the input parameters which we discuss below.
	
	At first sight, it appears from our potential and boundary condition that we have five free parameters,
	\begin{align}
	\left(\xi_{\text{IR}},(M_\star l)^3,\Phi_T,\Phi_P/\Phi_T,\epsilon\right) \ \ ,
	\end{align}
	and we plan to discuss each extensively.\footnote{When exploring the dependence on these parameters we ignore back-reaction constraints to streamline the discussion, but we ultimately take them into account in the following subsection.} The new parameter $\xi_{\text{IR}}$ is left to the next subsection and so, for now, we fix $\xi_{\text{IR}}=1$, restoring the usual scenario \cite{Kon2010,Ran2006,Cre2001,Nar2007}. In fact, solving the hierarchy problem imposes a constraint that the global minimum of the potential must lie at the TeV scale. This fixes the ratio $\Phi_P/\Phi_T$ as follows: the global extrema of the potential are located at 
	\begin{align}\label{extrema}
	\mu_{\pm} = k \left(\frac{\Phi_T}{\Phi_P}\right)^{1/\epsilon} \left[\frac{(4+\epsilon) \mp \sqrt{\epsilon(4+\epsilon)}}{2(2+\epsilon)}\right]^{1/\epsilon} \ \ ,
	\end{align}
	where, for $\epsilon>0$, $\mu_+$ and $\mu_-$ are the maximum and minimum respectively. Hence, once $\epsilon$ and $(M_\star l)$ are fixed\footnote{Throughout this section, a choice of $(M_\star l)$ determines $k$ by requiring the 4D Planck mass, Eq.~\eqref{4dplanck}, to be $10^{18} \text{GeV}$.}, the ratio $\Phi_P/\Phi_T$ is uniquely determined once we make our choice of $\mu_- = 1 \,\text{TeV}$.
	
	\begin{figure}[tbp]
	\center
  	\includegraphics[width=0.49\textwidth]{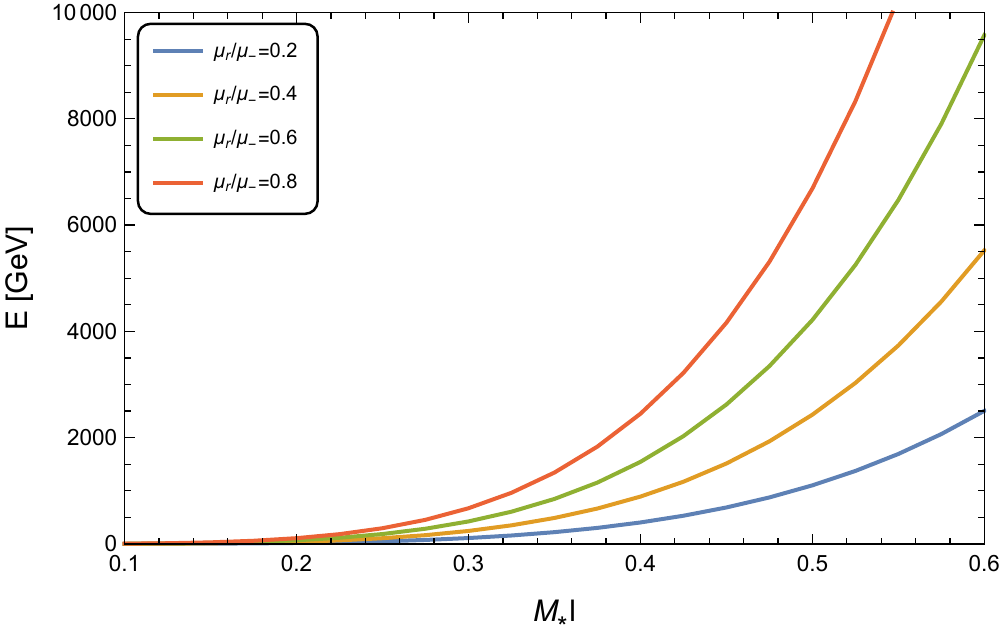} 
	\includegraphics[width=0.49\textwidth]{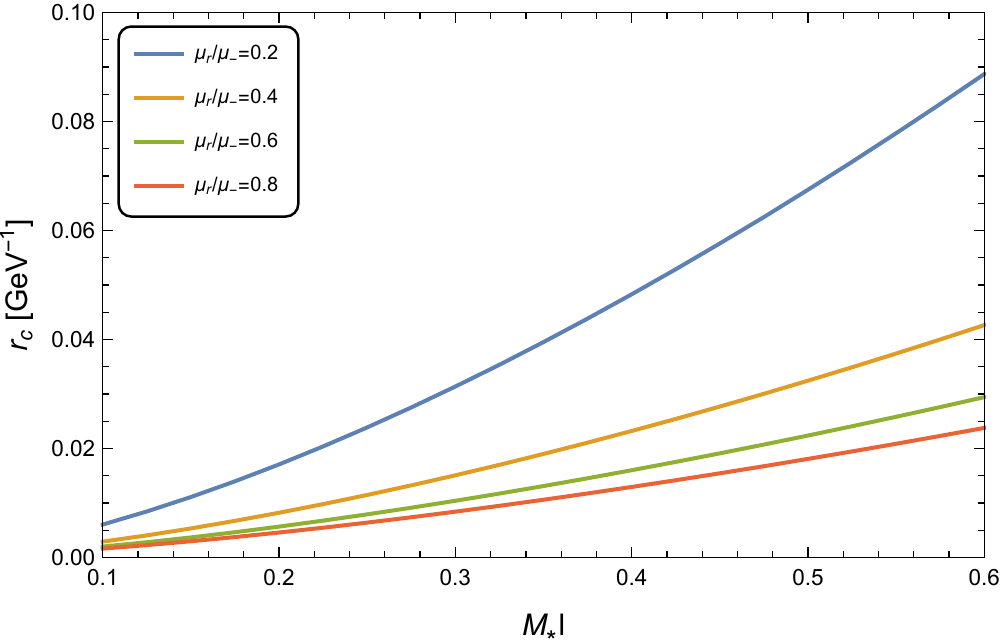} 
	\includegraphics[width=0.49\textwidth]{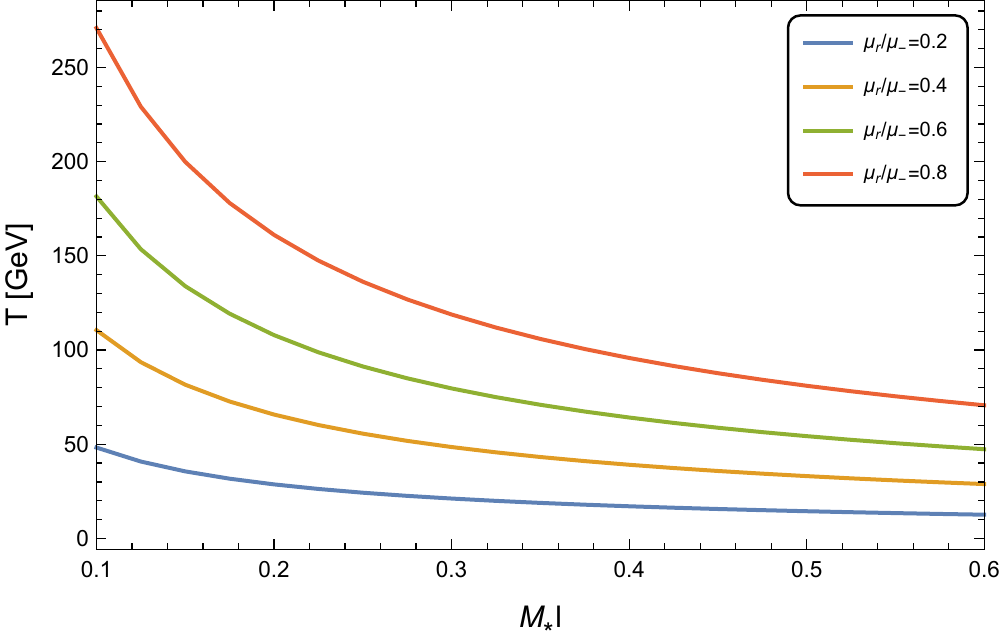} 
	\includegraphics[width=0.49\textwidth]{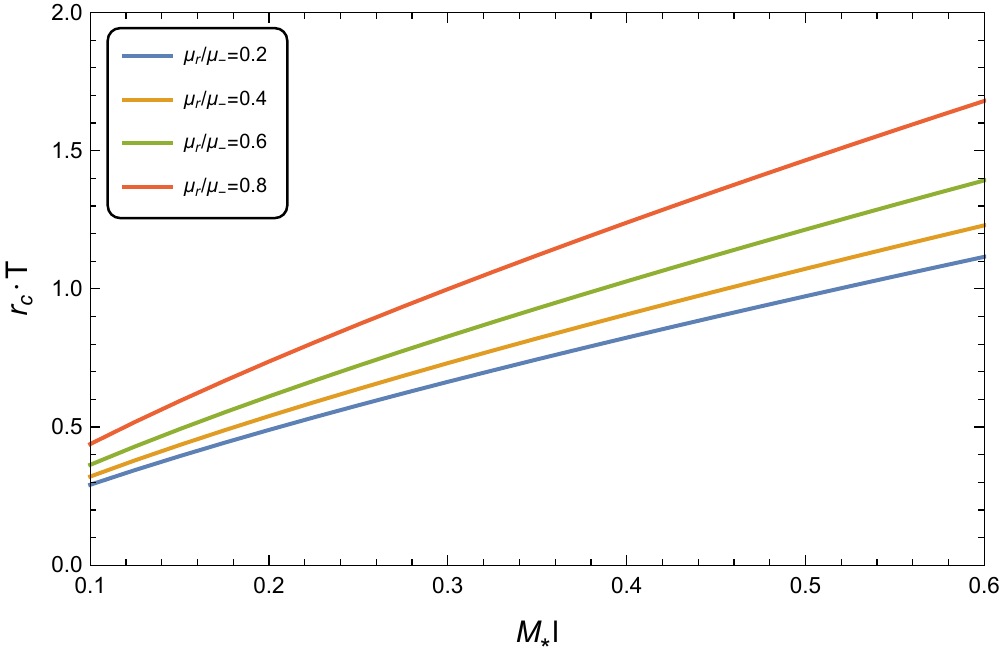}
  	\label{mlfigures}
	\caption{The dependence of various quantities in the bounce action $S_3/T$ of Eq.~\eqref{s3overTdecomp} on $(M_\star l)$ for a range of release points. We take the benchmark values $\epsilon = 0.05$ and $\Phi_T = 1$. \label{mlfigures}}
	\end{figure}
	
	\begin{figure}[tbp]
	\center
  	\includegraphics[scale=0.62]{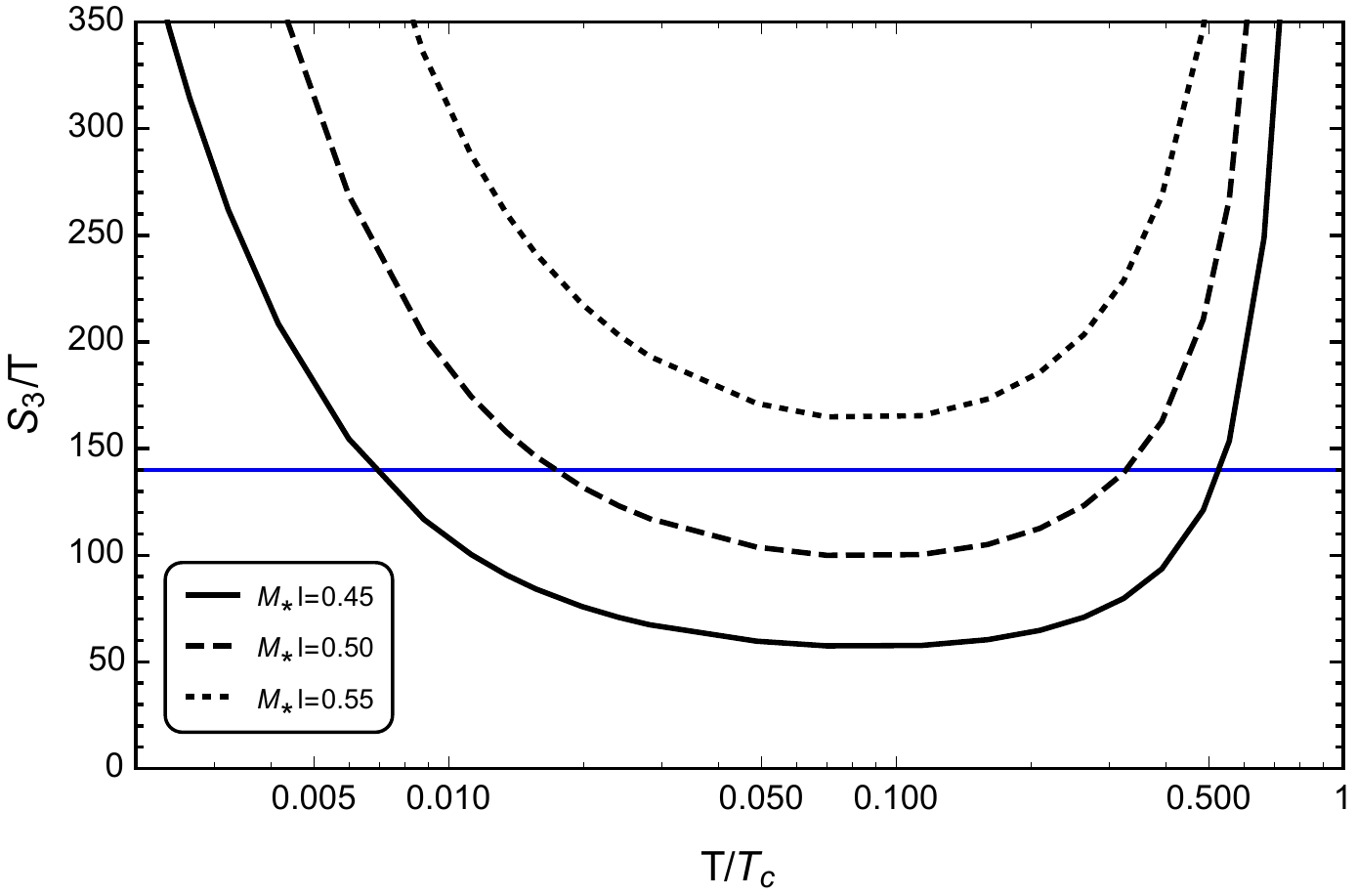}
  	\caption{The bounce action $S_3/T$ for various values of $(M_\star l)$ with the benchmark values $\epsilon = 0.05$ and $\Phi_T = 1$, where the blue line represents the nucleation condition. It is clear that increasing $(M_\star l)$ hinders the tunneling.\label{mls3overT}}
	\end{figure}
	
	\emph{$(M_\star l)$ factor.} The value of $(M_\star l)$ (corresponding to the number of colors in the CFT Eq.~\eqref{Ndef}) plays a crucial role in the phase transition. In Figure \ref{mlfigures}, the effect of $(M_\star l)$ on $E$, $T$ and $r_c$ is shown for various release points. Each factor displays a dependence on $(M_\star l)$ either explicitly, implicitly through $\mathcal{Z}$, or both. The rapid increase in the bubble energy, $E$, is due to both the $\mathcal{Z}$ factor and the increase in $r_c$. For $r_c$, the only dependence on ($M_\star l$) comes through $\mathcal{Z}$. Finally, the increase in $(M_\star l)$ for a fixed release point decreases the temperature. It is interesting to note that the combination $r_c T$ does not scale significantly with $(M_\star l)$. 	
	
	In Figure \ref{mls3overT}, we show the tunneling exponent for various values of $(M_\star l)$. Here, the devastating effect of large $(M_\star l)$ on the tunneling rate is clear. An incremental increase causes the system to get stuck in the false vacuum. This tension has been noted in Refs.~\cite{Kon2010,Ran2006,Cre2001,Nar2007} to be the most unpleasant feature of the phase transition. In fact, large $(M_\star l)$ (or $N$) is required for the semiclassical analysis to be reliable. We shall see below how the inclusion of $\xi_{\text{IR}}$ drastically changes the situation.
			
	\begin{figure}[tbp]
	\center
  	\includegraphics[width=0.49\textwidth]{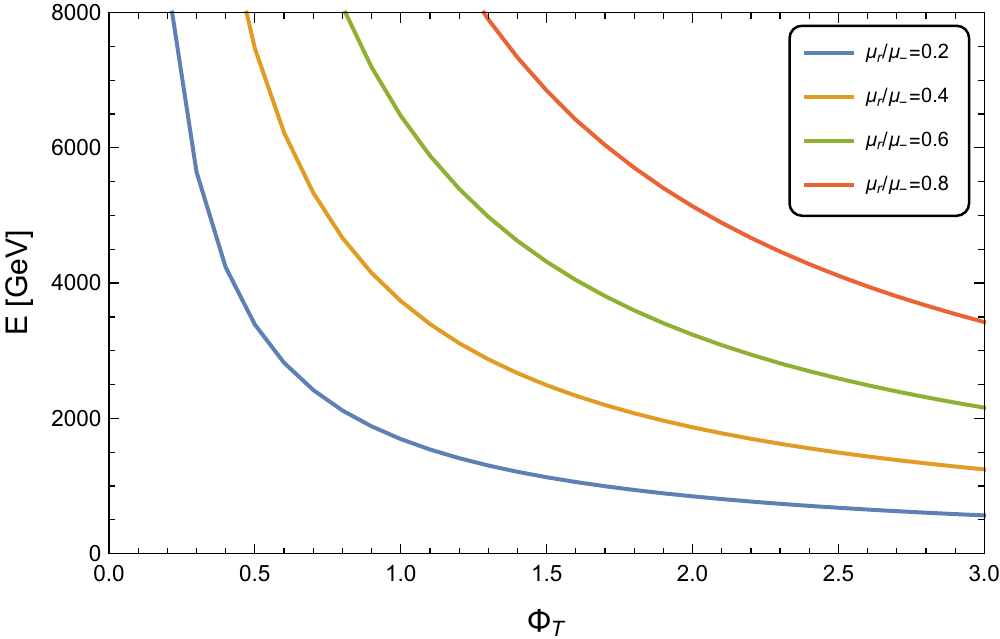} 
	\includegraphics[width=0.49\textwidth]{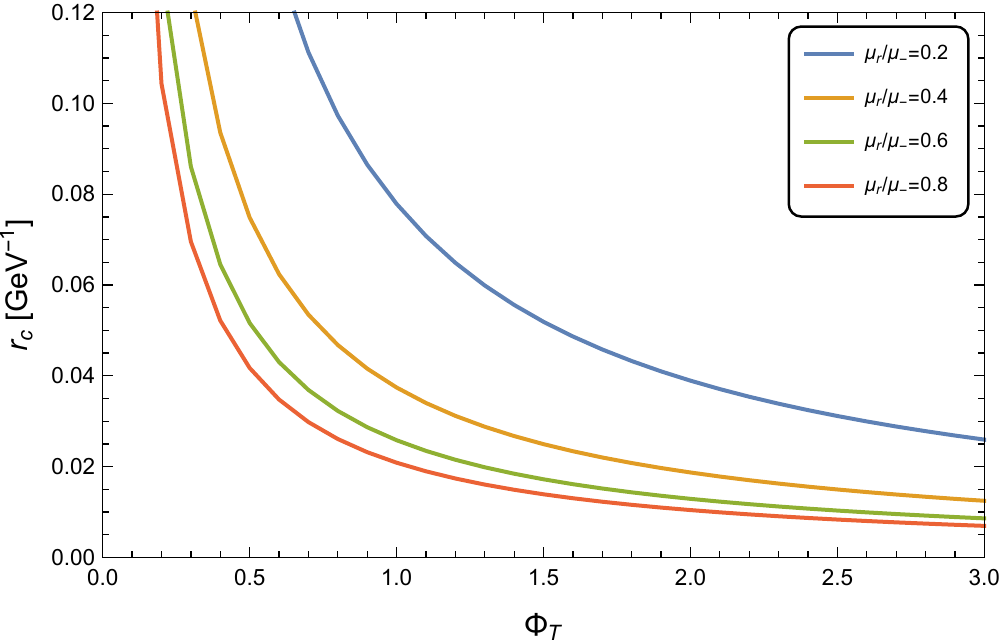} 
	\includegraphics[width=0.49\textwidth]{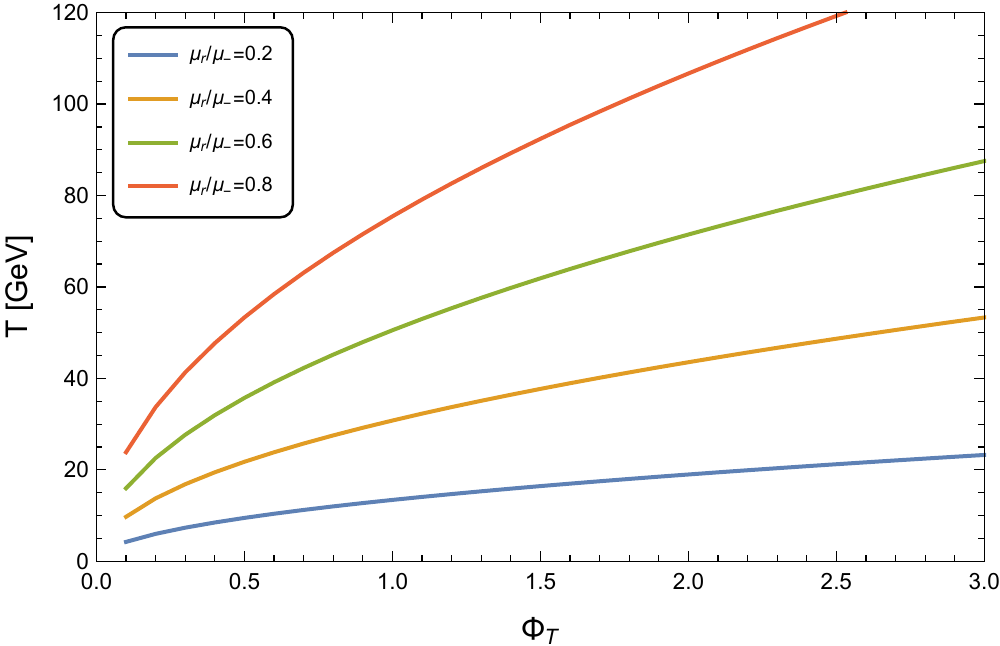} 
	\includegraphics[width=0.49\textwidth]{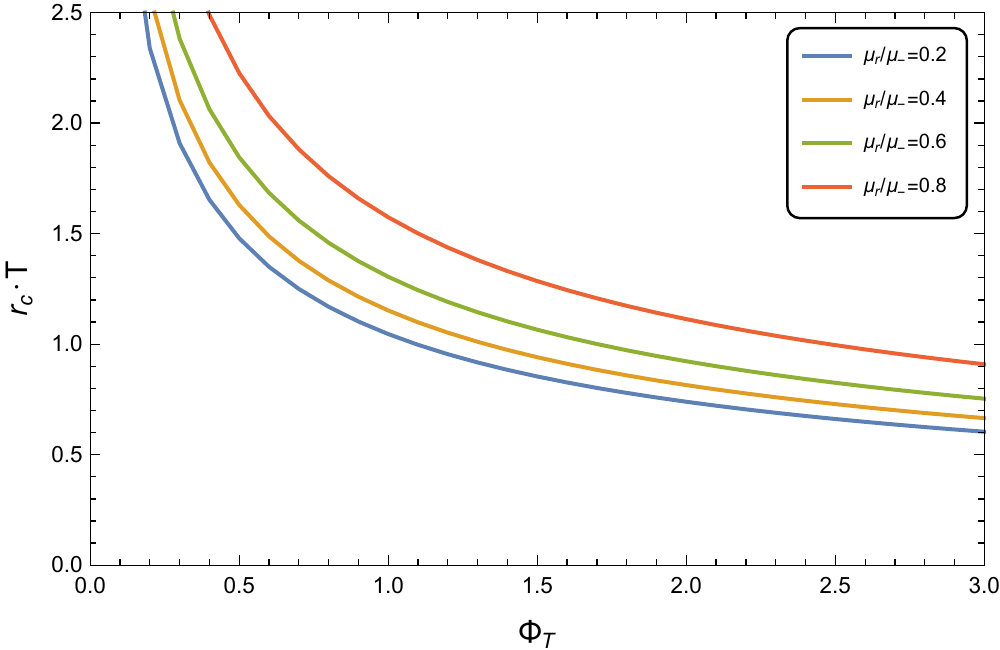}
	\caption{The dependence of various quantities in the bounce action $S_3/T$ of Eq.~\eqref{s3overTdecomp} on $\Phi_T$ for a range of release points. We take the benchmark values $\epsilon = 0.05$ and $(M_\star l) = 0.55$.\label{phifigures}}
	\end{figure}

	\begin{figure}[tbp]
	\center
  	\includegraphics[scale=0.63]{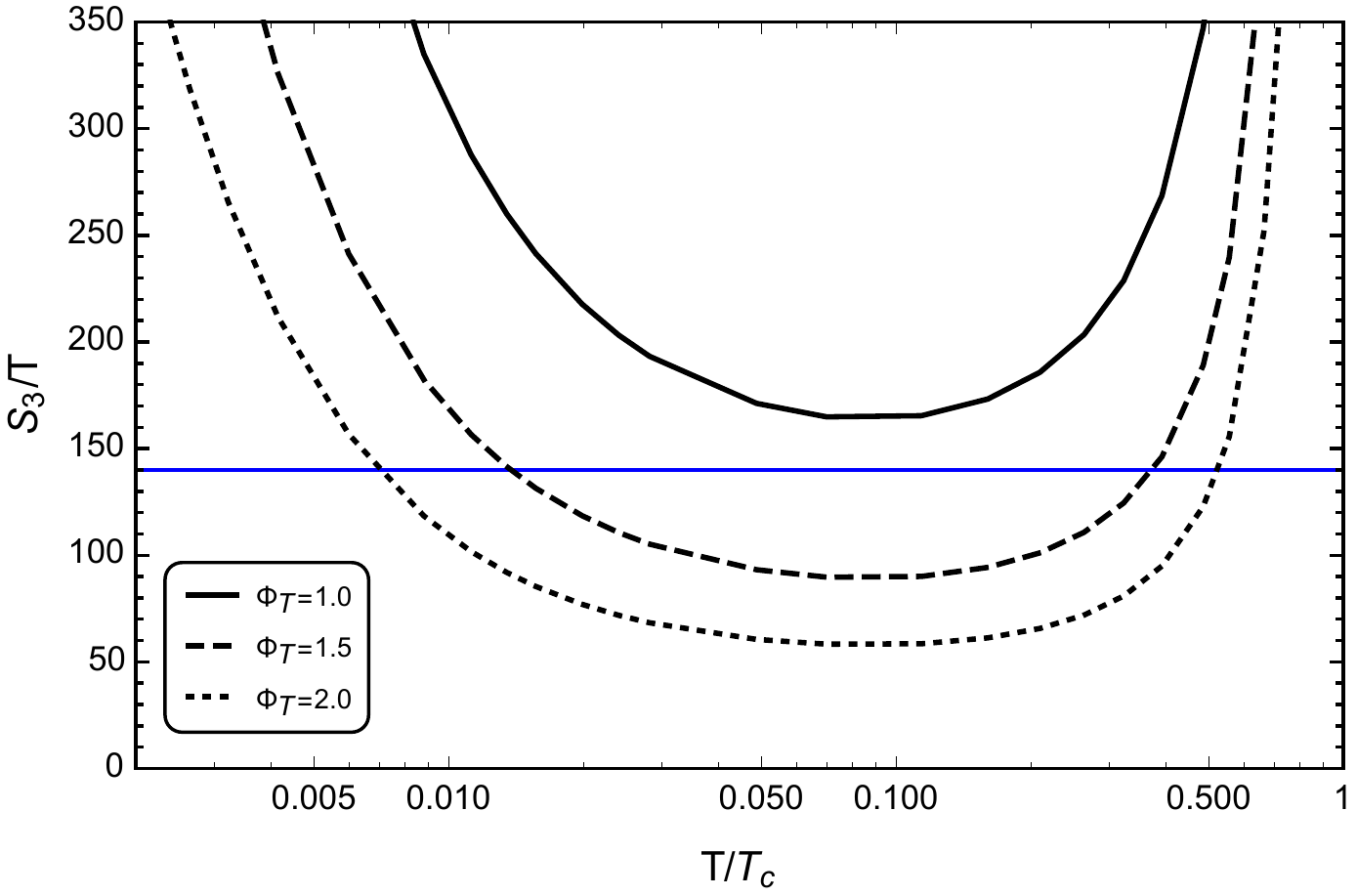}
  	\caption{The bounce action $S_3/T$ for various values of $\Phi_T$ with the benchmark values $\epsilon = 0.05$ and $(M_\star l)=0.55$, where the blue line represents the nucleation condition. It is clear that increasing $\Phi_T$ significantly improves nucleation rates.}
  	\label{phiTs3overT}
	\end{figure}	
					
	\emph{Effect of $\Phi_T$}. The parameter $\Phi_T$ plays a major role in determining the tunneling rate. Increasing $\Phi_T$ renders the potential deeper at the minimum, thereby considerably facilitating the tunneling to occur. We see in Figure \ref{phifigures} the effect of increasing $\Phi_T$ on the various quantities in Eq.~\eqref{s3overTdecomp}. As we shall explain below, with $\xi_{\text{IR}} = 1$ the condition of limited back-reaction on the background geometry prohibits us from leveraging the effect of $\Phi_T$. Arguably, the most important effect of brane-localized curvature will be to enable us to prop up $\Phi_T$ consistent with back-reaction constraints. Figure~\ref{phiTs3overT} shows the dramatic effect increasing $\Phi_T$ has on aiding nucleation.

	\begin{figure}[tbp]
	\center
  	\includegraphics[width=0.49\textwidth]{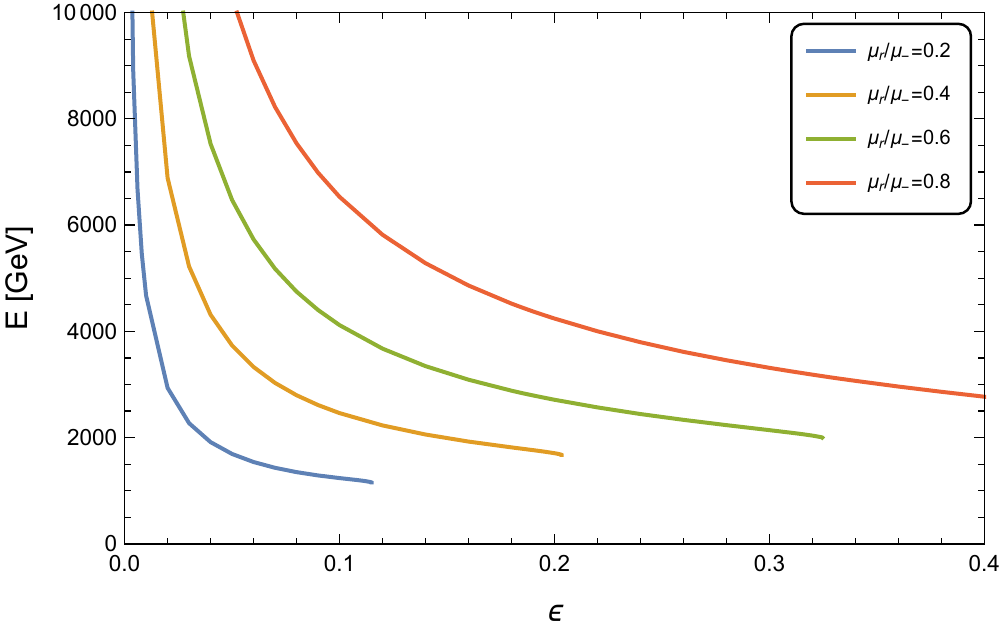} 
	\hfill
	\includegraphics[width=0.49\textwidth]{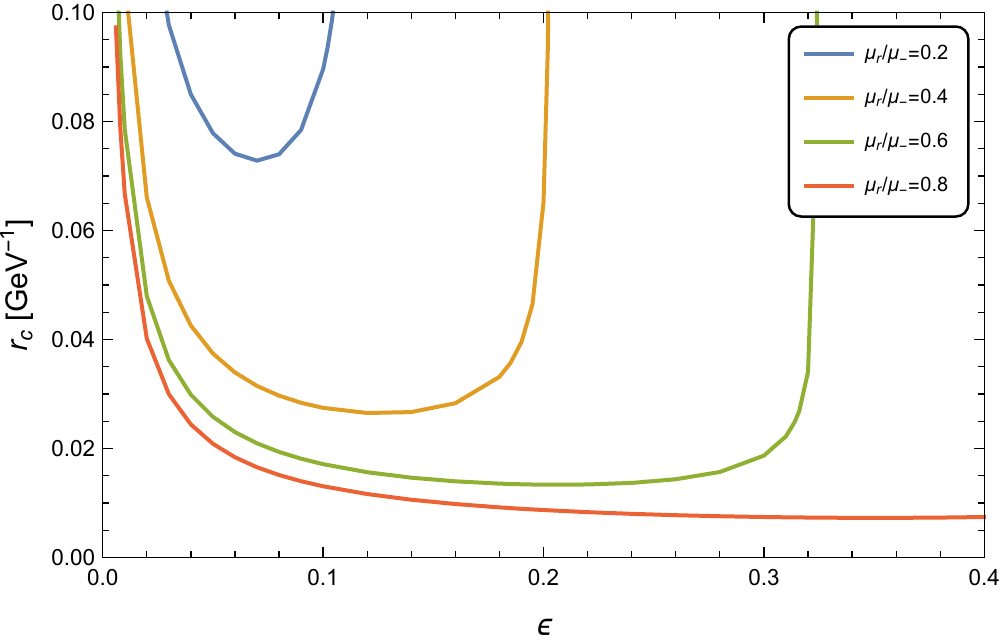} 
	\vfill
	\includegraphics[width=0.49\textwidth]{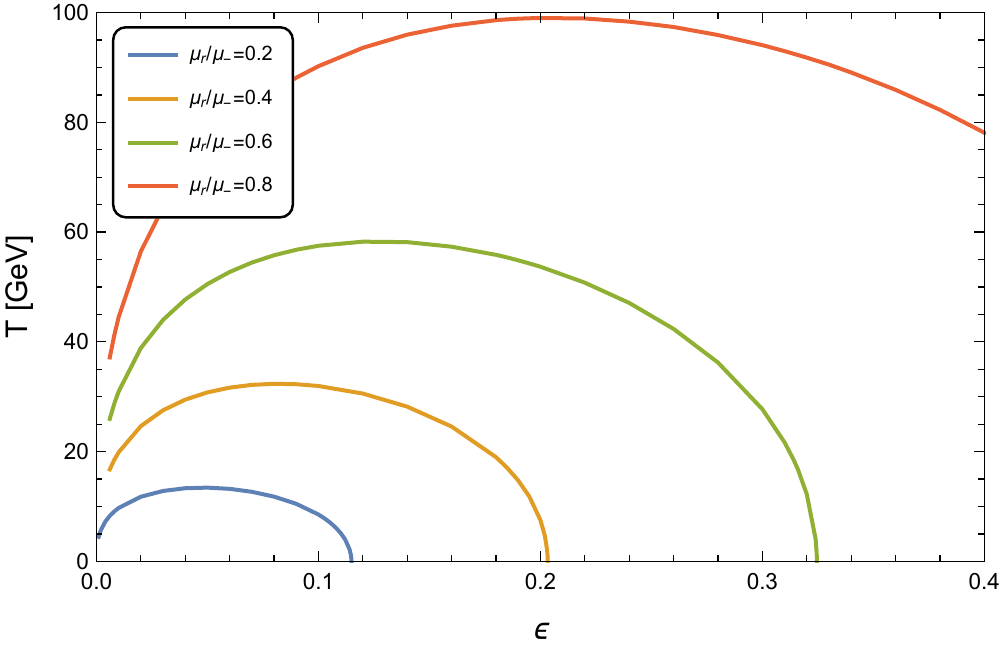} 
	\hfill
	\includegraphics[width=0.49\textwidth]{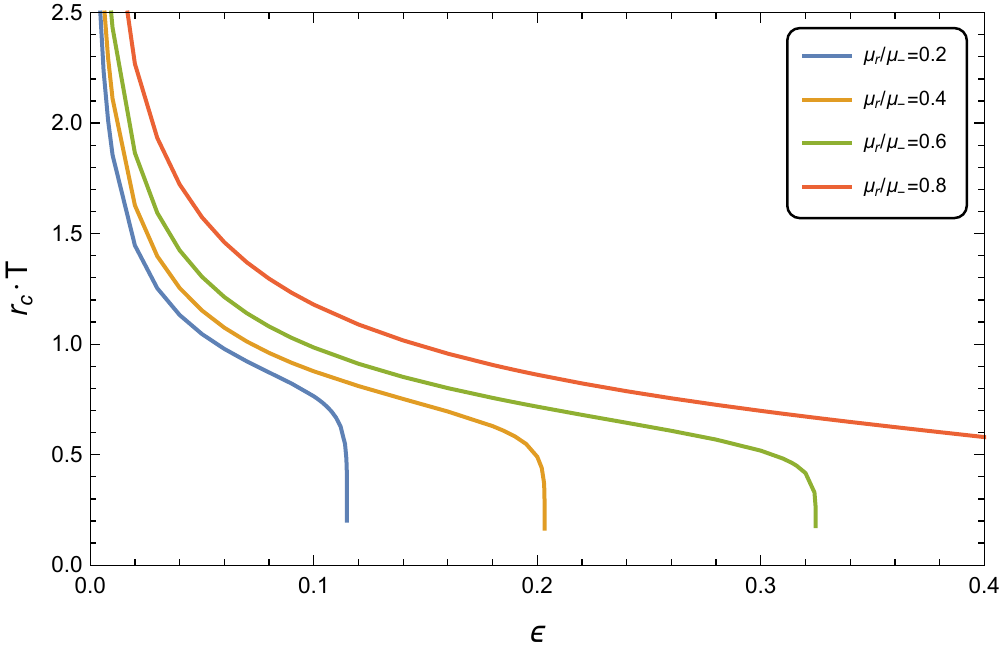}
  	\caption{The dependence of various quantities in the bounce action $S_3/T$ of Eq.~\eqref{s3overTdecomp} on $\epsilon$ for a range of release points. We take the benchmark values $\Phi_T = 1$ and $(M_\star l) = 0.55$.\label{epsilonfigures}}
	\end{figure}
		
	\begin{figure}[tbp]
	\center
  	\includegraphics[scale=0.62]{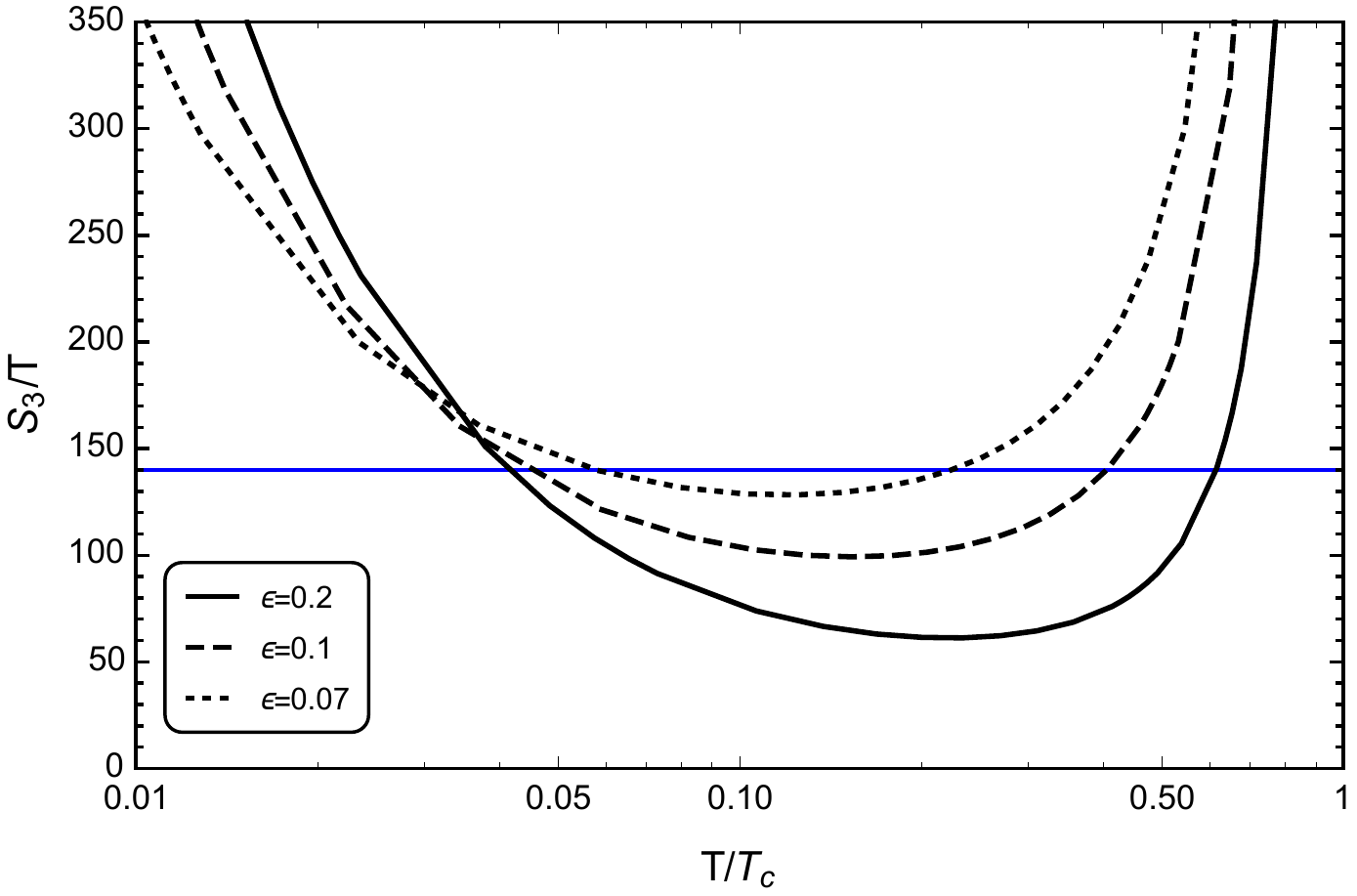}
  	\caption{The bounce action $S_3/T$ for various values of $\epsilon$ with the benchmark values $\Phi_T=1$ and $(M_\star l)=0.55$, where the blue line represents the nucleation condition. It is clear that, for positive values, decreasing $\epsilon$ increases the supercooling.}
  	\label{epss3overT}
	\end{figure}
	
	\emph{$\epsilon$ parameter}. The choice of $\epsilon$ has a strong impact on the strength of the phase transition. First of all, it uniquely determines the ratio between the extrema of the potential
	\begin{align}\label{min_max_ratio}
	\frac{\mu_-}{\mu_+} =\left(\frac{1+\sqrt{\epsilon/(4+\epsilon)}}{1-\sqrt{\epsilon/(4+\epsilon)}}\right)^{1/\epsilon} \ \ ,
	\end{align}
	and thus a larger $\epsilon$ pushes $\mu_+$ closer to the minimum. This indeed facilitates the tunneling taking place, as seen in Figure~\ref{epss3overT}. Conversely, one could attain substantial amounts of supercooling by lowering $\epsilon$. Another important factor is the barrier height compared to the depth of the potential
	\begin{align}\label{barrier_height_ratio}
	\left| \frac{V(\mu_+)}{V(\mu_-)} \right| = \left(\frac{\mu_+}{\mu_-}\right)^4 \ \ ,
	\end{align}
	which is exact thanks to the properties of the potential. It is clear that a larger $\epsilon$ leads to a larger barrier with respect to the depth of the minimum, thus hindering the tunneling from proceeding. Hence, the size of $\epsilon$ presents two competing effects on the phase transition. 
	
	Figure~\ref{epsilonfigures} shows the effect $\epsilon$ has on the various quantities in the bounce action of Eq.~\eqref{s3overTdecomp}. We observe rather more complex behaviors compared to the monotonic dependences on $(M_\star l)$ and $\Phi_T$. The bubble energy decreases with increasing $\epsilon$ as $\mu_+$ is brought closer to the minimum, until the barrier height becomes so large that the field can no longer reach $\mu=0$. The situation for $r_c$ and $T$ is more interesting. Here, it is adequate to think in terms of a single particle moving in the inverted potential of Eq.~\eqref{radionpotoffset}. Evidently, the size of $\epsilon$ controls the {\em flatness} of the potential between the extrema. Increasing $\epsilon$ introduces a larger gradient in the potential, i.e. $\partial V/\partial\mu$ becomes large between the extrema. Thus, the particle rolls faster leading $r_c$ to decrease. Eventually, the minimum of the inverted potential becomes so deep that it becomes harder for the particle to climb out of the well. This is the reason why $r_c$ begins to increase beyond a certain $\epsilon$. The same physics explains the behavior of the temperature.

	\subsection{Turning on $\xi_{\text{IR}}$}\label{turningonxi}
	
	Turning on $\xi_{\text{IR}}$, the dynamics of the phase transition changes dramatically. On the one hand, having $\xi_{\text{IR}} \ll 1$ significantly enhances the tunneling rate by suppressing the energy of the bubble. On the other hand, the size of $\Phi_T$ can be made larger thus prompting the phase transition to quickly proceed. To clarify the latter point, let us recall that in order to ignore back-reaction of the bulk scalar on the background geometry, the following constraints must be satisfied~\cite{Gol1999}
	\begin{align}\label{limitedBR}
	\Phi_T \ll (M_\star l)^{3/2}, \quad \Phi_P \ll (M_\star l)^{3/2} \ \ .
	\end{align}
	 
	 If $\xi_{\text{IR}} =1 $, an enormous constraint on the parameter space is imposed since a very small $(M_\star l)$ is required to have a small $\mathcal{Z}$, allowing the phase transition to proceed. The role of $\xi_{\text{IR}}$ is crucial in ameliorating this tension: $(M_\star l)$ could be substantially large while the normalization factor $\mathcal{Z}$ stays $\mathcal{O}(1)$. In particular, it is now possible to increase $\Phi_T$, consistent with Eq.~\eqref{limitedBR}, to expedite the phase transition. 
	
\begin{figure}[tbp]
	\center
  	\includegraphics[width=0.49\textwidth]{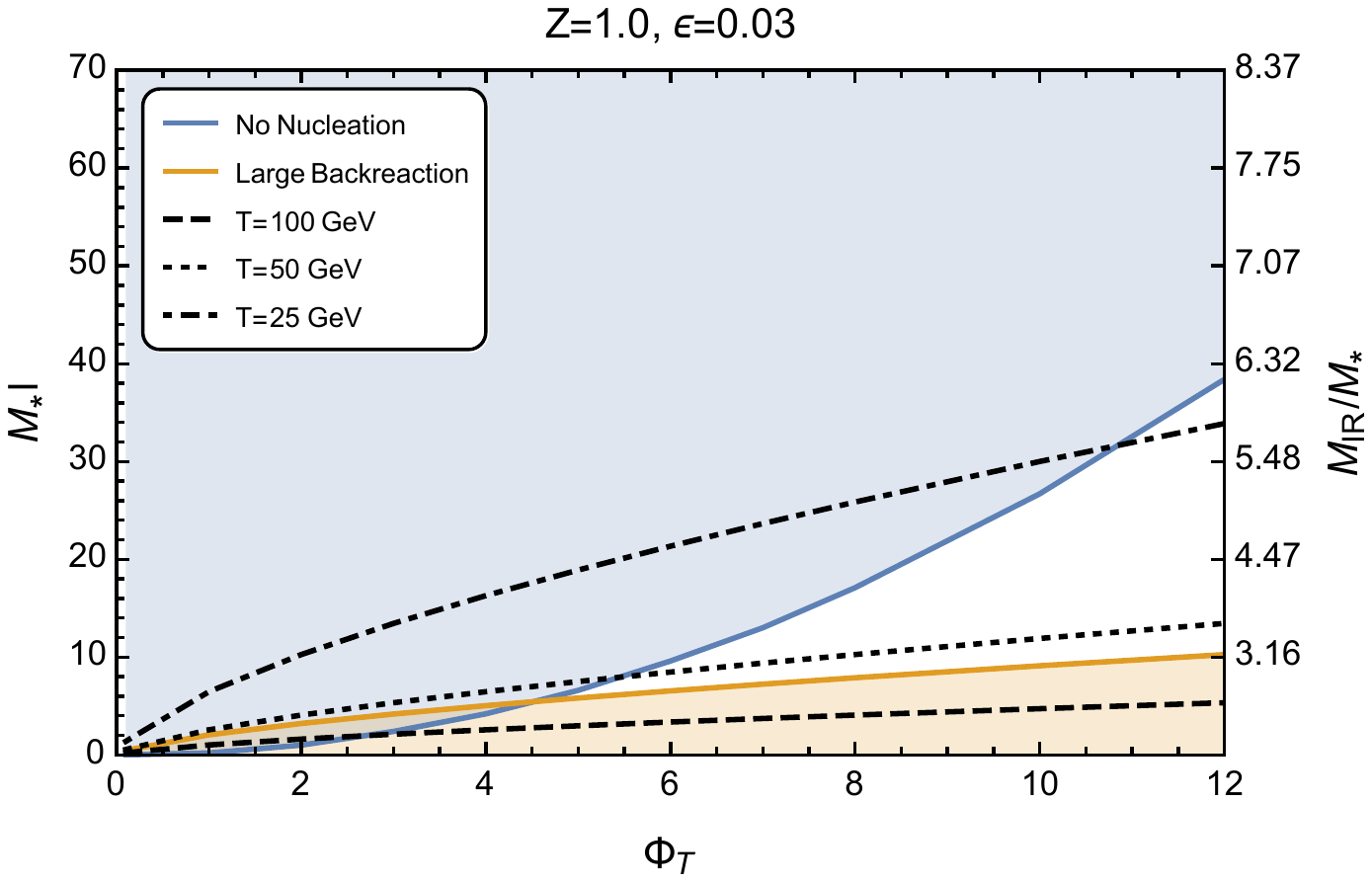} 
	\hfill
	\includegraphics[width=0.49\textwidth]{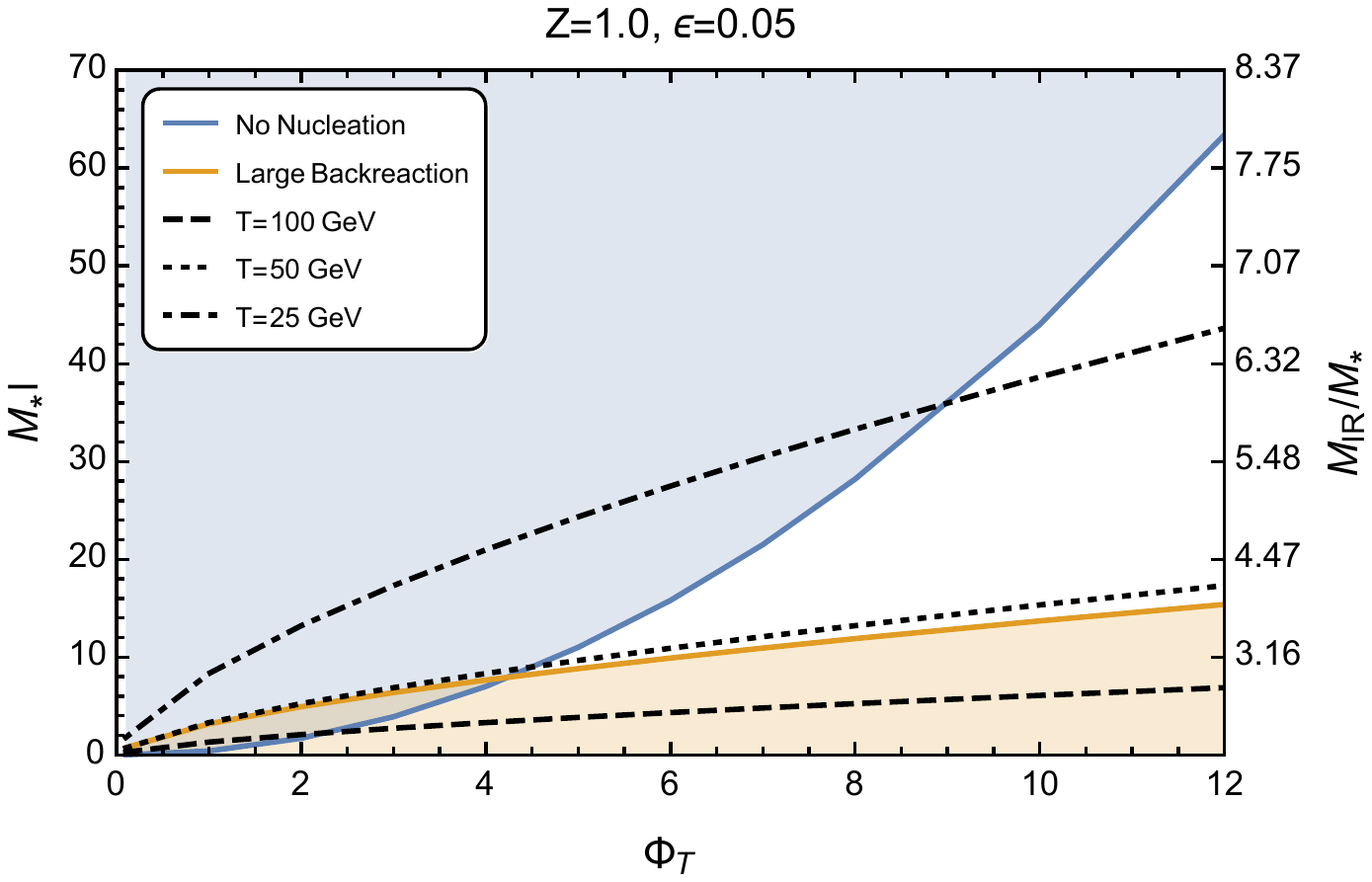} 
	\vfill
	\includegraphics[width=0.49\textwidth]{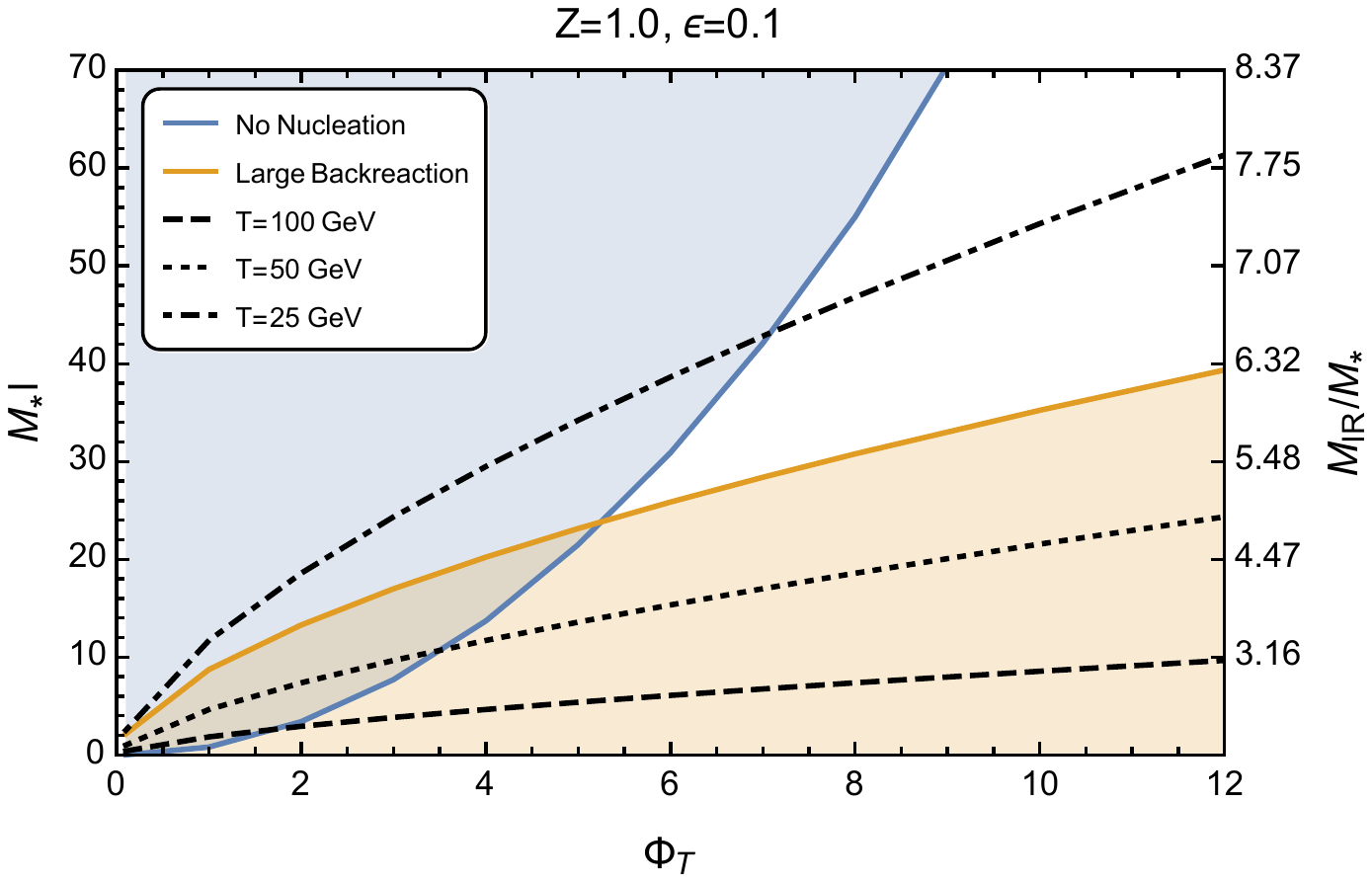} 
	\hfill
	\includegraphics[width=0.49\textwidth]{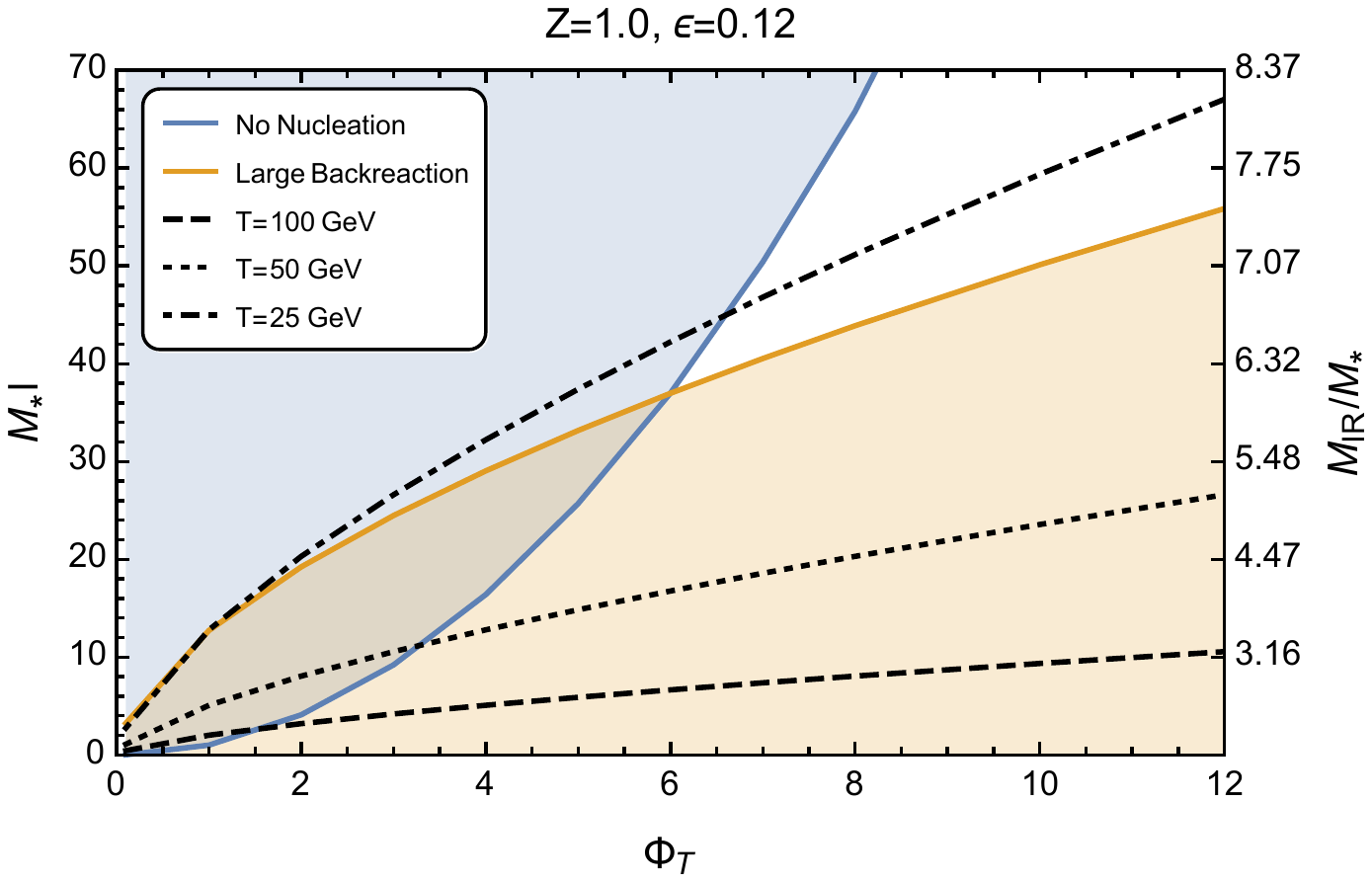}
	\vfill
	\includegraphics[width=0.49\textwidth]{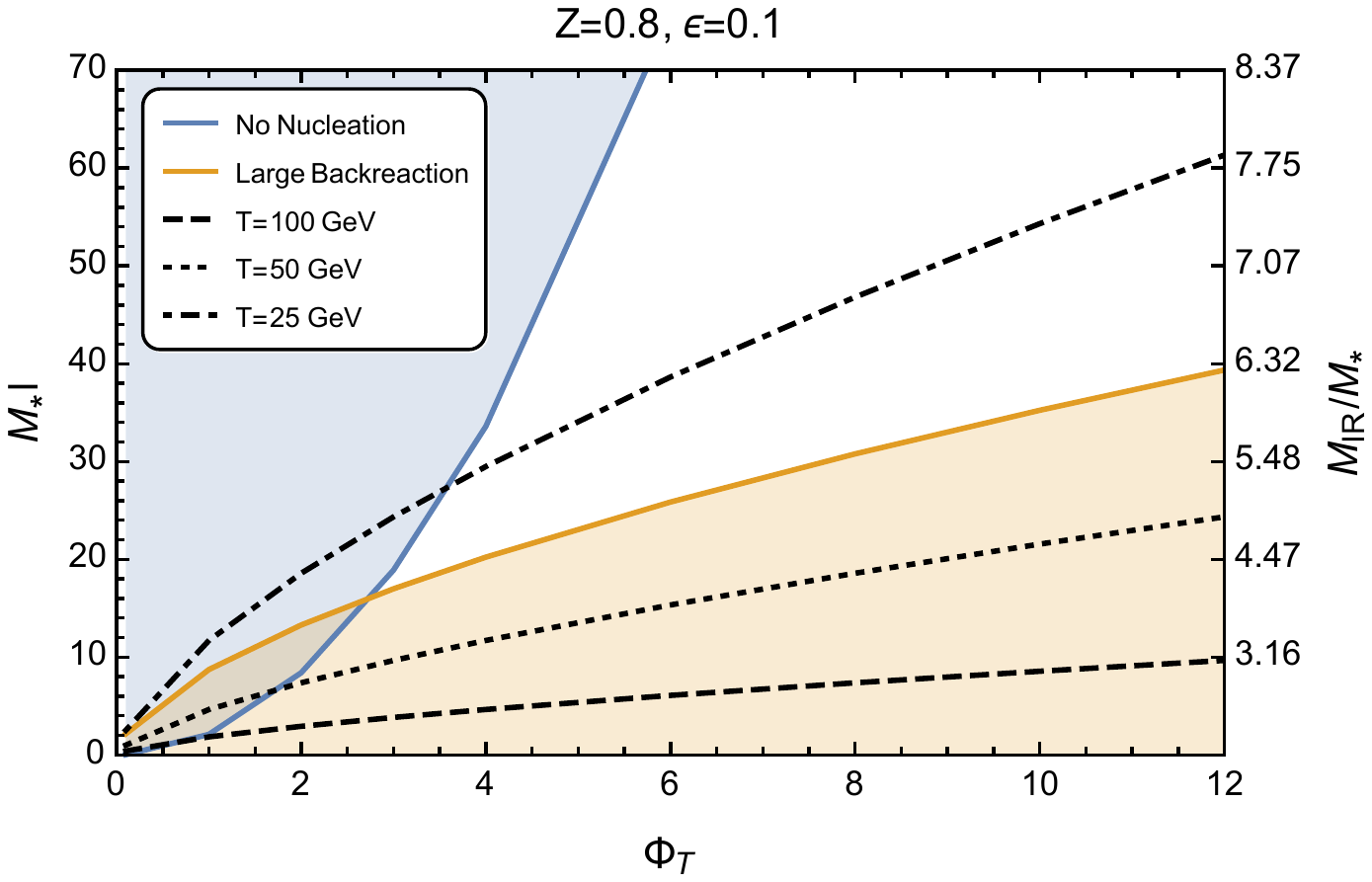}
	\hfill
	\includegraphics[width=0.49\textwidth]{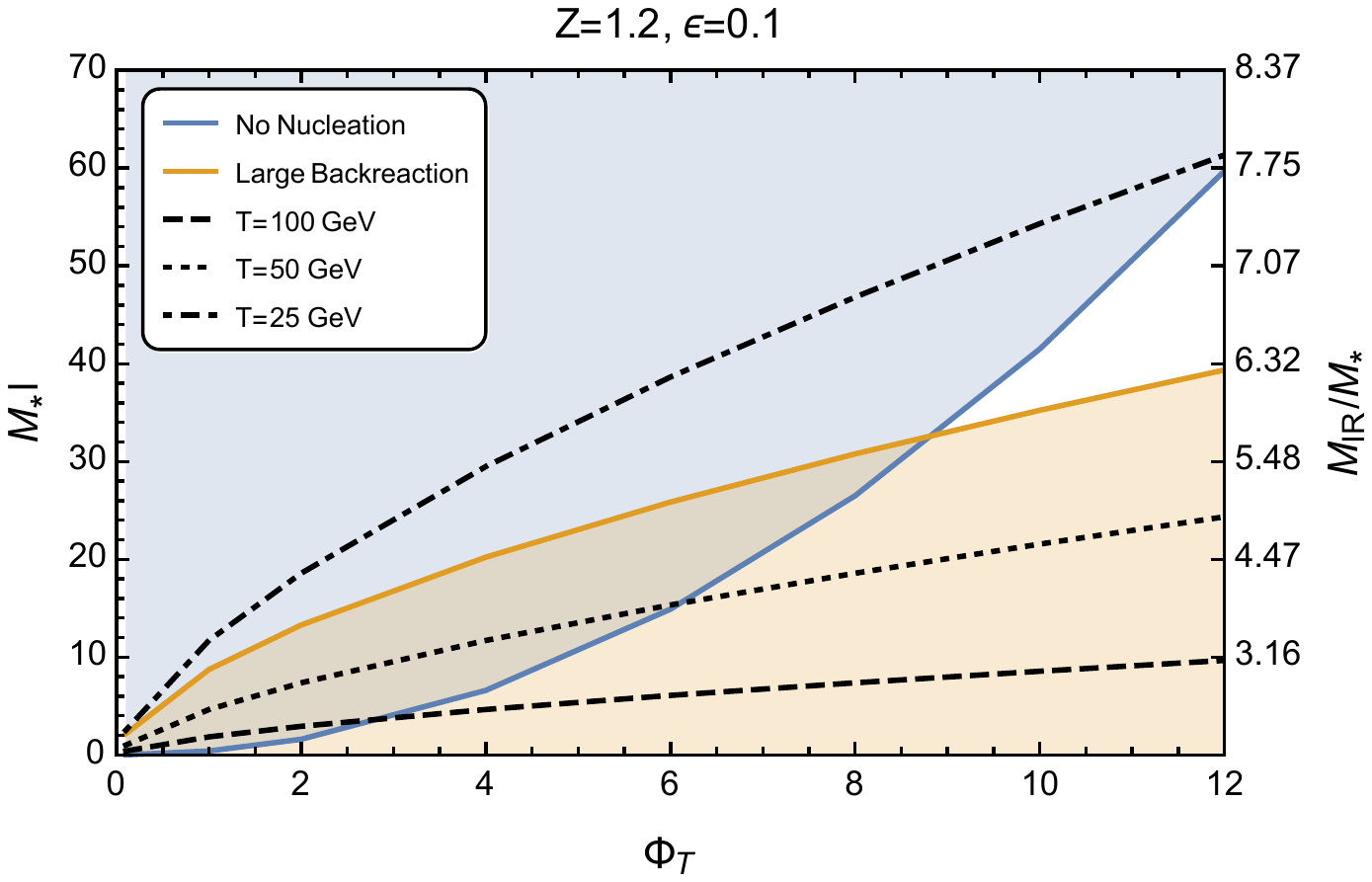}
  	\caption{The parameter space of $(M_\star l)$ and $\Phi_T$ for varying $\epsilon$ and $\mathcal{Z}$ values. The blue-shaded region represents values at which nucleation is never achieved while the orange-shaded region represents values at which back-reaction is no longer negligible. The backreaction limit is found by setting $\Phi_P=(M_\star l)^{3/2}$ in Eq.~\eqref{extrema} and solving for $(M_\star l)$. On the right axis we show the ratio of the fundamental mass scales $M_{\text{IR}}/M_\star$ found from Eq.~\eqref{znorm}.}
  	\label{parspacez1}
	\end{figure}

	We begin by finding the $(M_\star l)-\Phi_T$ parameter space for various $\epsilon$ and $\mathcal{Z}$ values, shown in Figure~\ref{parspacez1}. We remind the reader that $\Phi_P$ is fixed by the requirement that $\mu_- = 1 \, \text{TeV}$. On one hand, nucleation does not take place for parameters in the blue-shaded region. On the other, the parameters in the orange-shaded region induce large back-reaction. This latter region is found by setting $\Phi_P=(M_\star l)^{3/2}$ in Eq.~\eqref{extrema} and solving for $(M_\star l)$ as a function of $\Phi_T$, given that $\mu_- = 1$ TeV. It is always true that $\Phi_P>\Phi_T$, and hence $\Phi_P$ serves as the greater threat to the small back-reaction approximation. On the right axis, we include the ratio of the fundamental mass scales $M_{\text{IR}}/M_\star$ found from Eq.~\eqref{znorm}. Notice this ratio has a weak dependence on $\mathcal{Z}$, and so for the range of values in Figure~\ref{parspacez1} we observe no significant change.
	
	The sole effect of $\mathcal{Z}$ on the parameter space is changing the size of the blue-shaded region. Smaller $\mathcal{Z}$ significantly enhances the tunneling probability, thus shrinking the disallowed region in the parameter space. Only a $20\%$ decrease drastically enlarges the allowed parameter space as we clearly see in Figure~\ref{parspacez1}. This behavior is quite non-trivial since this $20\%$ decrease, for example, corresponds to an infinitesimal change in the ratio of $M_{\text{IR}}/M_\star$. Indeed it is important to note that a certain level of fine-tuning is present in our approach since we always choose $M_\text{IR}/M_\star$ to keep $\mathcal{Z}$ fixed and $\mathcal{O}(1)$. If we fix $(M_\star l)$ and change the ratio $M_\text{IR}/M_\star$ by $1\%$ the normalization factor $\mathcal{Z}$ changes by $\mathcal{O}(100\%)$ or so. Extrapolating the behaviour shown in Figure~\ref{parspacez1}, the completion of the phase transition in this case would require substantially large values of both $\Phi_T$ and $(M_\star l)$.

	The role of $\epsilon$ is interesting but proceeds as in previous studies. A larger $\epsilon$ induces faster nucleation as evident from our previous discussion and Figure~\ref{parspacez1}. Yet, the orange-shaded region becomes larger with increasing $\epsilon$ as is clear from\footnote{With $\xi_{\text{IR}} = 1$, some previous papers \cite{Kon2010,Ran2006,Cre2001,Nar2007} have employed a brane tension term in the potential of the form $\delta T_1\mu^4$ to allow for a negative $\epsilon$. In particular, such a scenario relaxes back-reaction limits since $\Phi_P<\Phi_T$, which permits nucleation at $(M_\star l)$ of $\mathcal{O}(1)$.} Eq.~\eqref{extrema}. Constant-$T_c$ curves are plotted using Eq.~\eqref{Tcrit}. Perhaps the most important effect of $\epsilon$ is to enable one to accommodate higher values of $T_c$, which may be crucial for some applications. For example, by decreasing $\epsilon$ one could attain higher $T_c$ values. Notice, however, that one cannot continue pushing $\epsilon$ to lower values and expect to achieve larger $T_c$. Below some {\em optimum} $\epsilon$, the decrease in constant-$T_c$ curves begins to outpace that of the orange-shaded region, hindering the system from reaching higher $T_c$ values. 	
	
	\begin{figure}[tbp]
	\center
  	\includegraphics[width=0.49\textwidth]{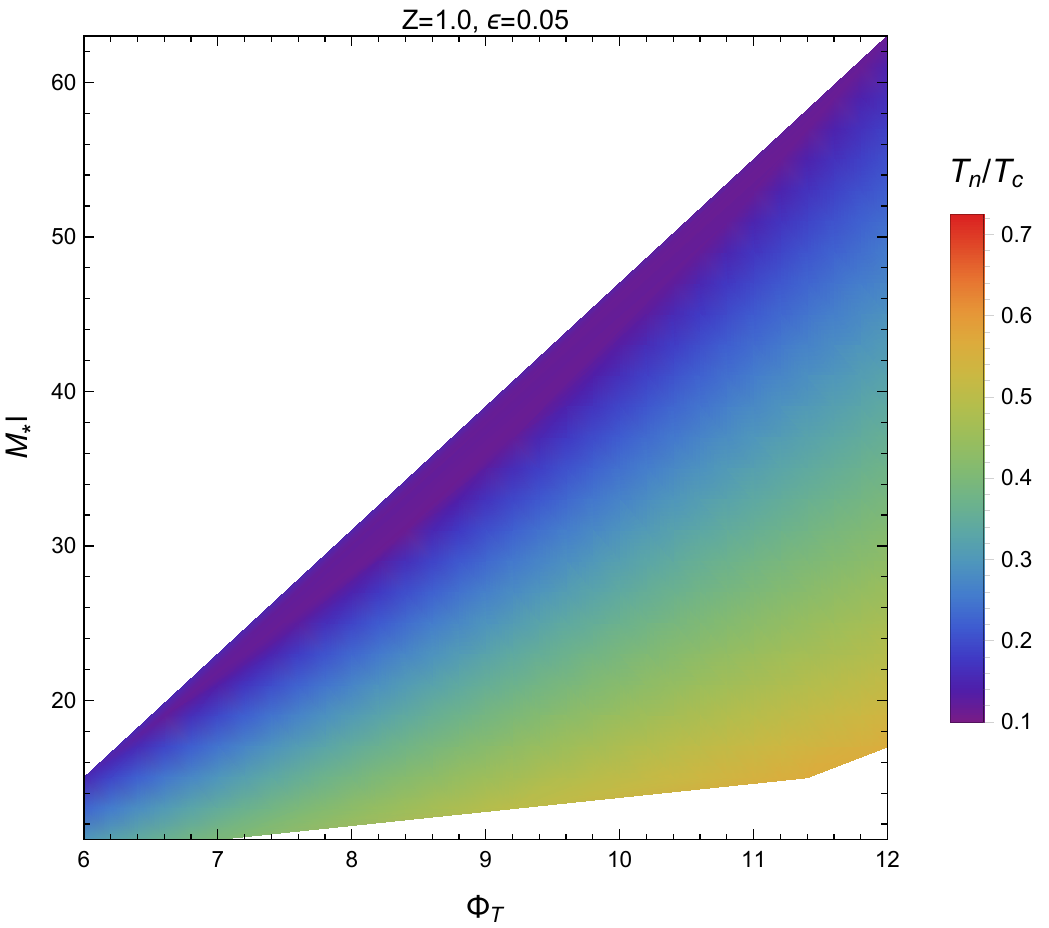} 
	\hfill
	\includegraphics[width=0.49\textwidth]{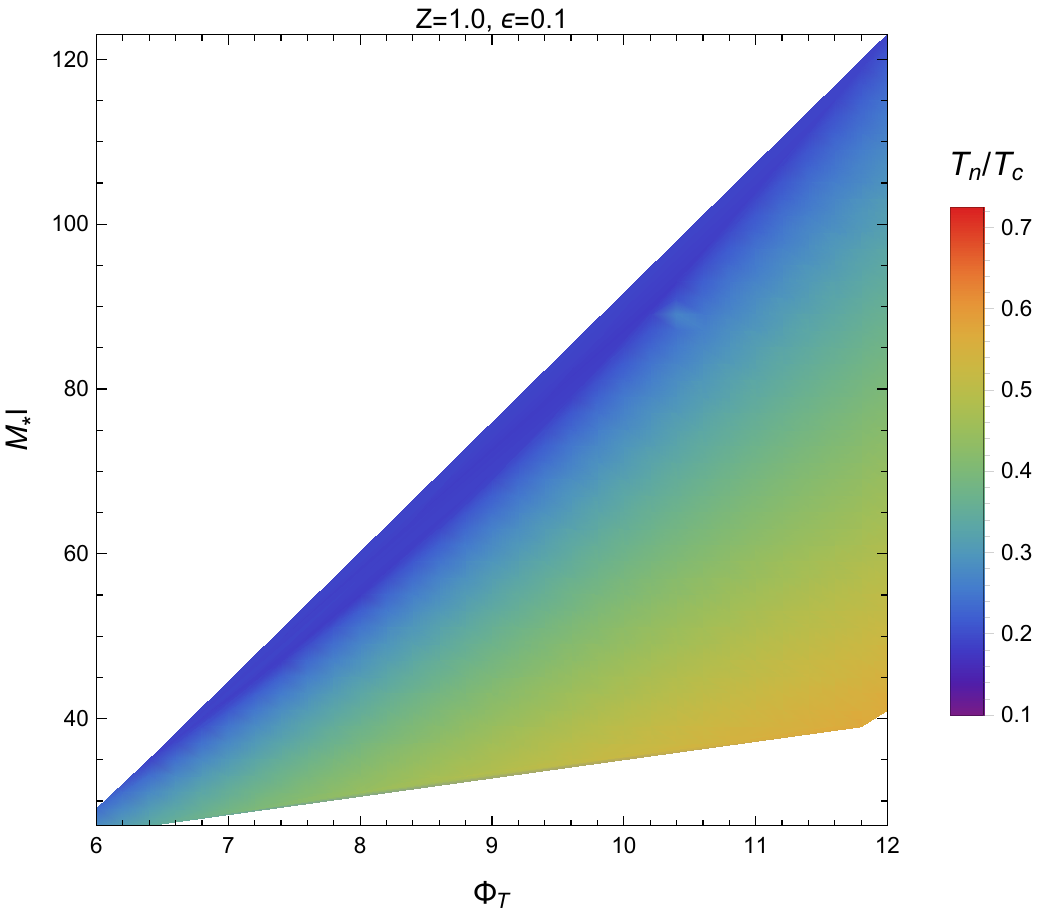} 
	\vfill
	\includegraphics[width=0.49\textwidth]{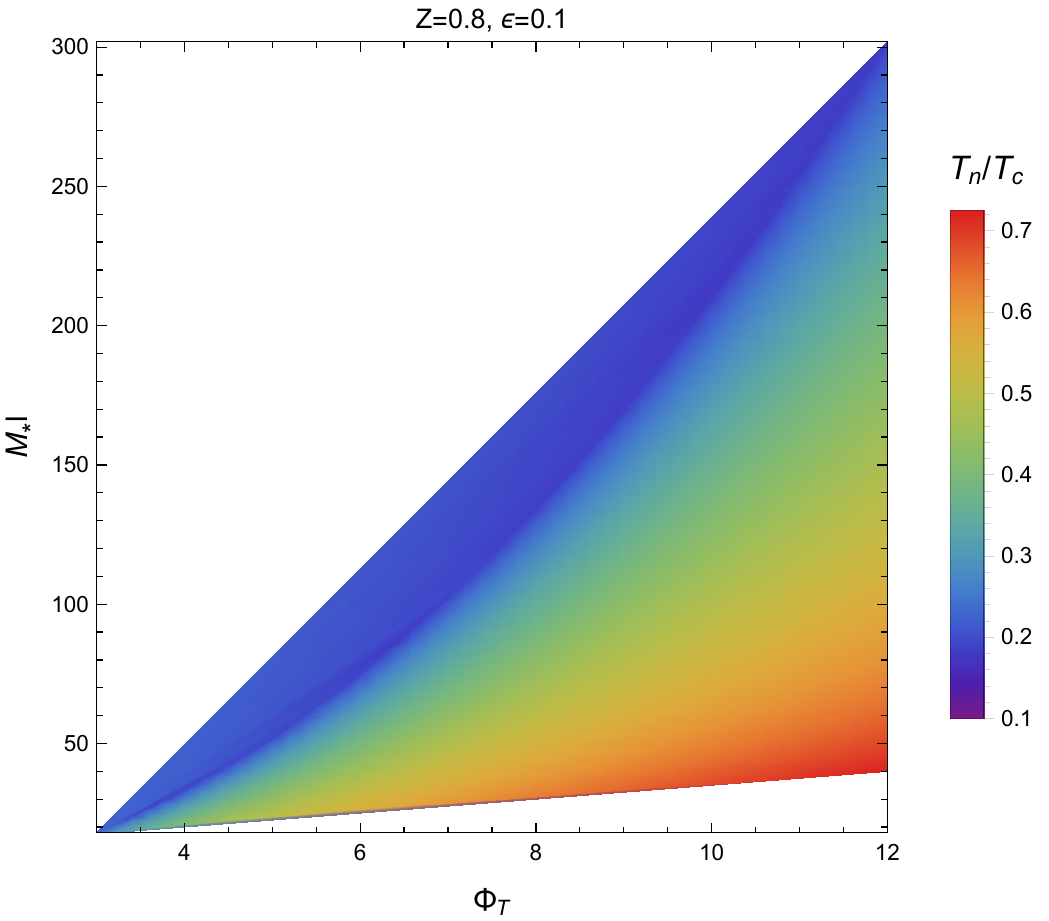} 
	\hfill
	\includegraphics[width=0.49\textwidth]{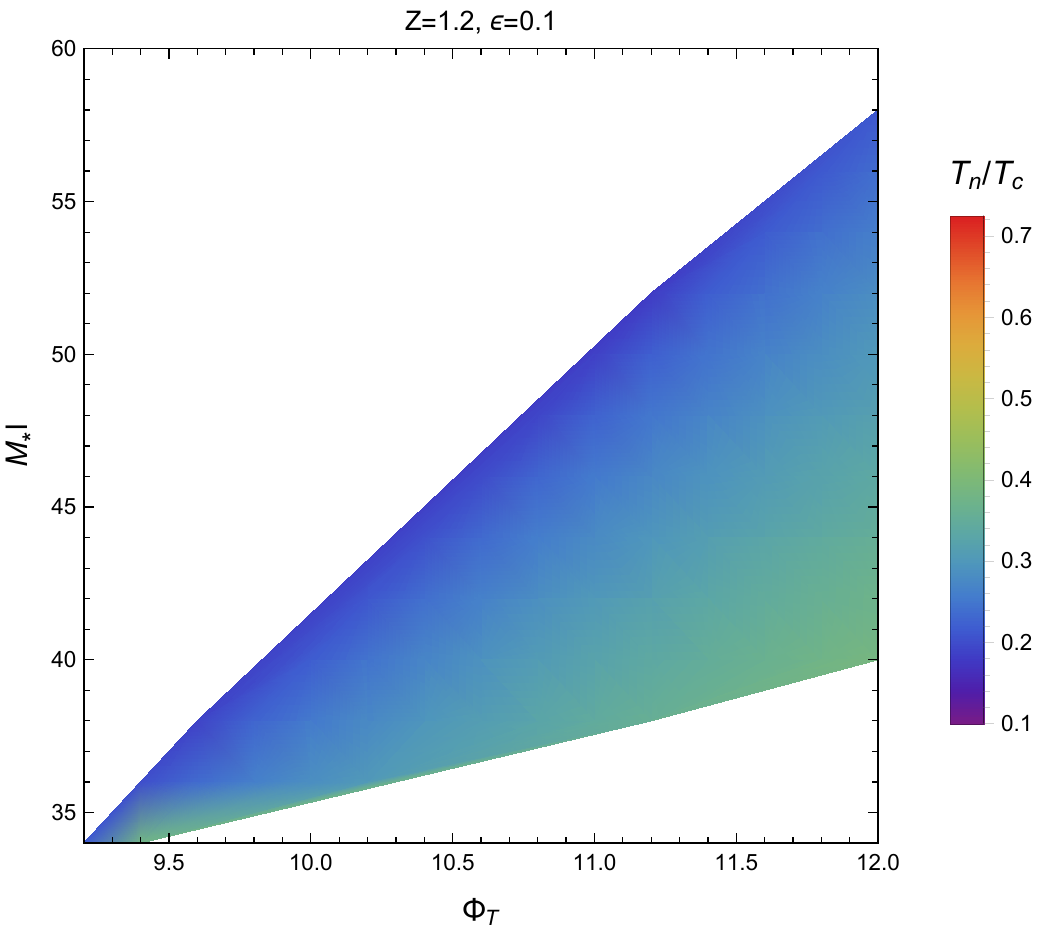} 
  	\caption{The variation in supercooling $T_n/T_c$ over the allowed parameter regions for a selection of the parameter space plots in Figure~\ref{parspacez1}.\label{densityplots}}
	\end{figure}
	
\begin{figure}[tbp]
	\center
  	\includegraphics[width=0.49\textwidth]{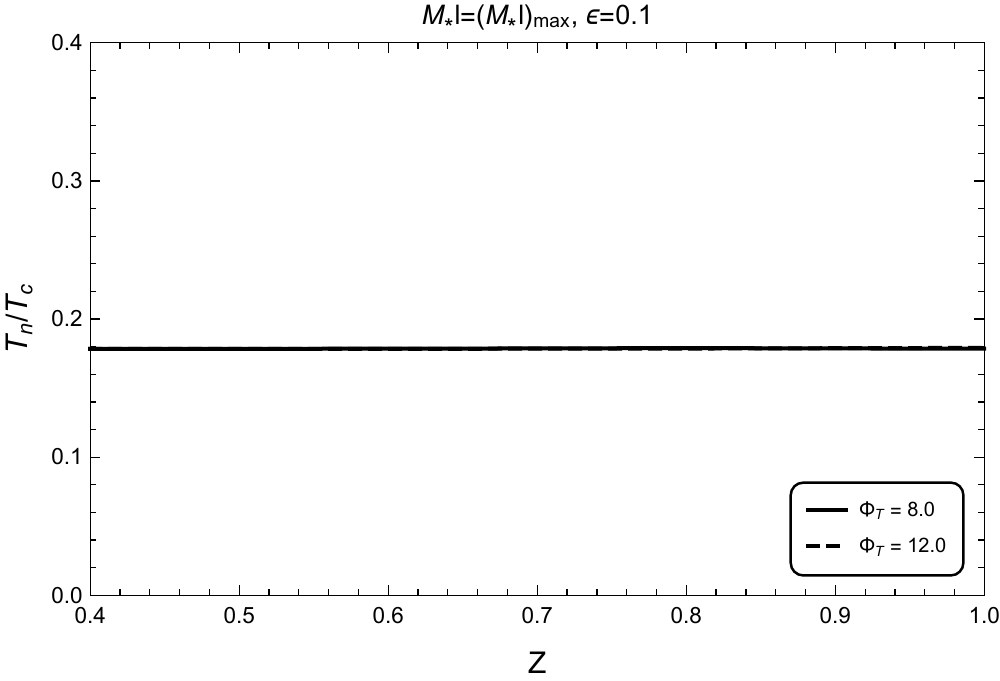} 
	\hfill
	\includegraphics[width=0.49\textwidth]{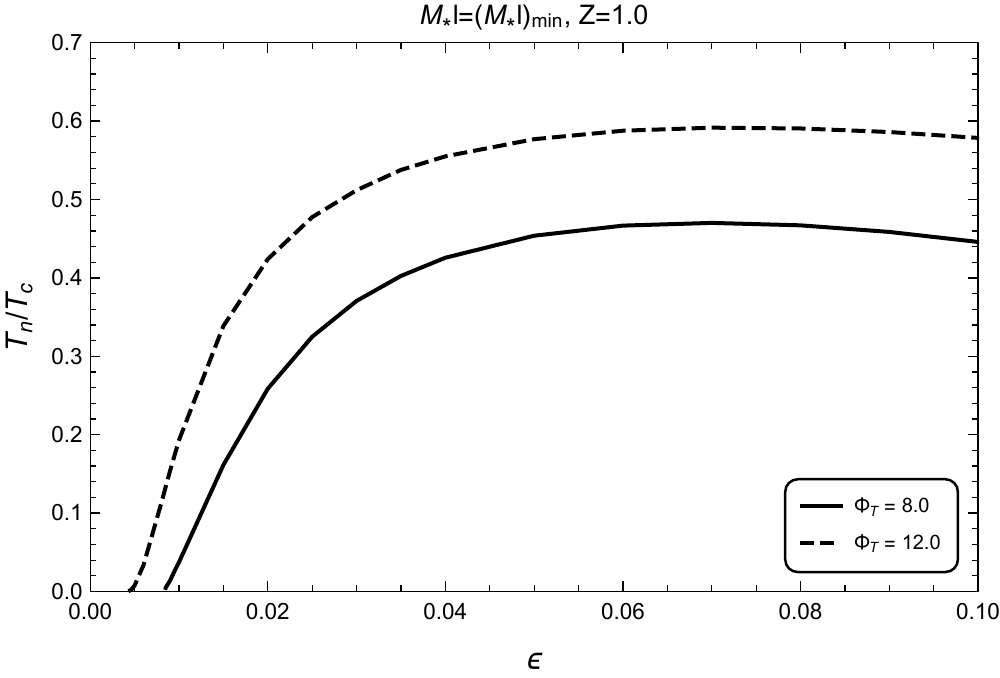} 
  	\caption{Left: The variation in the minimum $T_n/T_c$ (which is found at the maximum $(M_\star l)$) with respect to $\mathcal{Z}$ for $\epsilon=0.1$ and $\Phi_T=8.0,\,12.0$. Notice the two $\Phi_T$ lines are degenerate. Right: The variation in the maximum $T_n/T_c$ (which is found at the minimum $(M_\star l)$) with respect to $\epsilon$ for $\mathcal{Z}=1.0$ and $\Phi_T=8.0,\,12.0$.\label{tntcvszmax}}
	\end{figure}

	We now ask ourselves the question, what is the level of supercooling throughout our available parameter space? Figure~\ref{densityplots} shows color-maps for some of the allowed parameter regions in which red represents larger $T_n/T_c$ (less supercooling) while blue represents smaller $T_n/T_c$ (more supercooling). We immediately observe that our parameter regions allow for $T_n/T_c\sim \mathcal{O}(0.1-1)$, a significant difference to the previous results of Refs.~\cite{Kon2010,Nar2007} in which large amounts of supercooling were unavoidable. The next question is how low a $T_n/T_c$ could we accommodate within the allowed parameter space. It is quite interesting that $\mathcal{Z}$ has almost no effect in this regard as shown in the left plot of Figure~\ref{tntcvszmax}. On the contrary, the right plot of Figure~\ref{tntcvszmax} shows that lowering $\epsilon$ allows for significantly smaller $T_n/T_c$. Notice that the right plot uses the lowest possible $(M_\star l)$, which implies there is more room to attain large amounts of supercooling.
	
	To summarize, brane-localized curvature allows the phase transition to be completed at much larger values of $(M_\star l)$. In particular, this implies that the holographic description is well motivated and the semi-classical treatment is well suited to describe the physics. Such large values of $(M_\star l)$ imply large values of the brane kinetic term,\footnote{It is clear from our plots that lowering $M_{\text{IR}}$, while keeping $\mathcal{Z}$ fixed, does not allow the PT to take place. Indeed, in the extreme limit $M_{\text{IR}} \to 0$, and as previous works have shown, one has to consider quite small values of $M_\star l$ and $\Phi_T$. It is not possible to visualize in our plots this tiny corner of the parameter space, but we have indeed checked that our numerical routines reproduce the results in the literature when we take $M_{\text{IR}}$ to vanish identically.} as the parameter space discussed in this work typically requires $M_{\text{IR}}/M_\star \sim \mathcal{O}(1)-\mathcal{O}(10)$, see Figure~\ref{parspacez1}. One should inquire how natural our set-up is.
First of all, the inclusion of brane-localized curvature is necessary in 5D models \cite{Dva2000,Carena:2001xy}. It suffices to notice that massive matter confined to the brane generates a divergent contribution proportional to Eq.~(\ref{branecurv}). This is rather simple to understand if we realize that $M_{\text{IR}}$ looks like the Planck mass in ordinary 4D gravity, and the latter receives well-known quantum corrections from massive matter (see, for example, Ref.~\cite{Bir1982}). Second, since our operator is not forbidden by any symmetries in the model, it is natural to explore renormalized values that satisfy the range of parameters we explored. Brane kinetic terms in 5D models have also been studied in the context of gauge fields, and similar conclusions were reached \cite{Carena:2002me}. Finally, the presence of a brane kinetic term for gravity does not directly affect the stabilization via the Goldberger-Wise (GW) mechanism. Nevertheless, it has been shown that the presence of such terms contributes to a Casimir energy in the 5D bulk, which provides an alternative stabilization mechanism for 5D models \cite{Ponton:2001hq}. In the future, we hope to explore alternative stabilization mechanisms within the context of our model.

\subsection{Alternative look at the black hole contribution}
It has been argued in Ref.~\cite{Cre2001} that the black hole contribution to the tunneling exponent scales as $N^2$. The upshot of our model is to considerably enhance, at large $N$, the radion tunneling by decreasing its kinetic energy. Here, one might generally worry that at large $N$ the black hole contribution might kick in and dominate the tunneling process, upsetting some of the successes of our model. Indeed, in the absence of the exact instanton configuration, one cannot assess with certainty the dynamics of the high-temperature phase. Nevertheless, in the original analysis of Ref.~\cite{Cre2001}, it was suggested to model the black hole by a field variable that describes the position of the horizon and investigate the tunneling problem by stitching together the free energies of both phases.

\begin{figure}[tbp]
	\center
  	\includegraphics[scale=0.7]{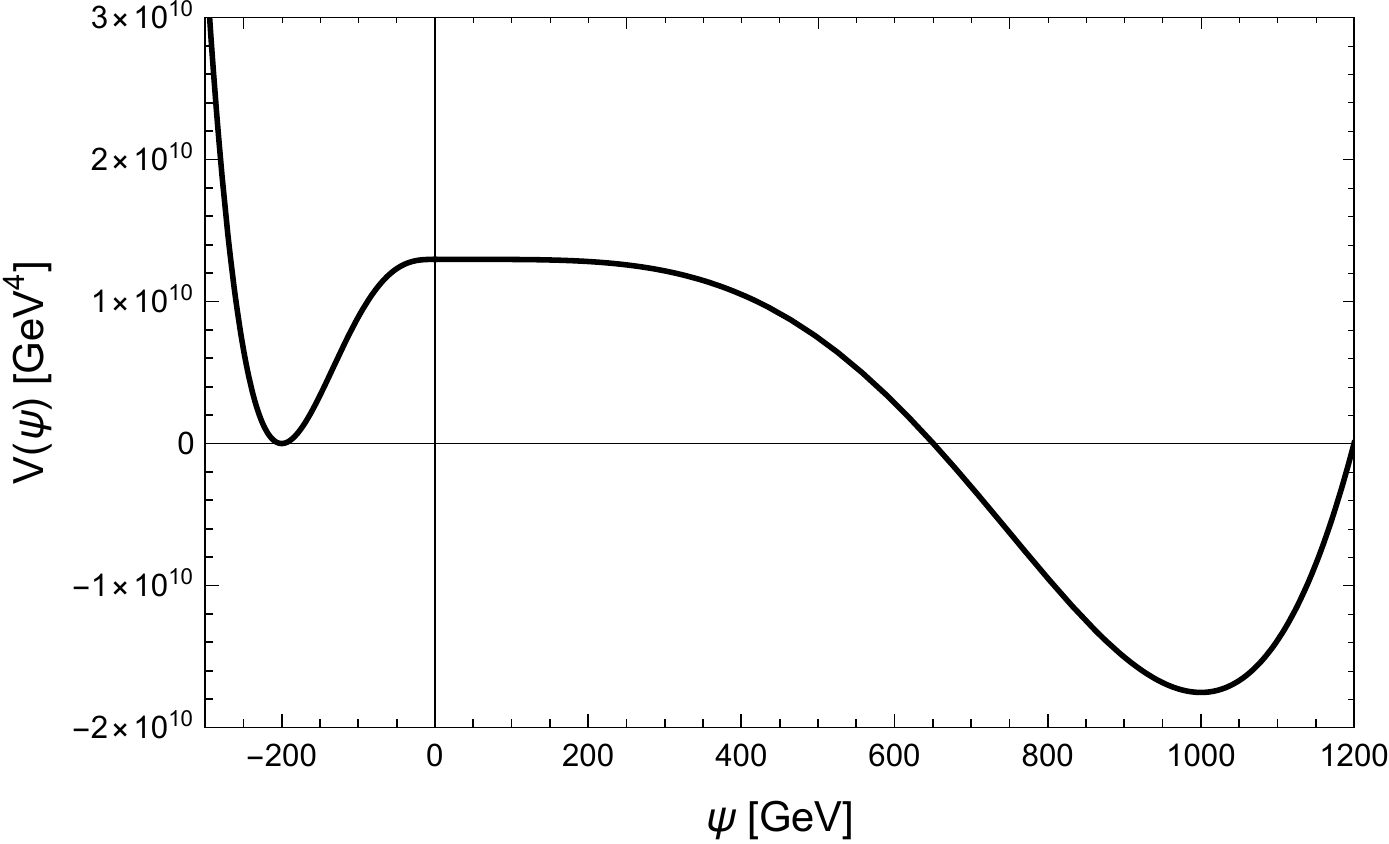}
  	\caption{A plot of the combined potential, described by Eq.~\eqref{comb_pot}, for $\Phi_T=1,\ \epsilon=0.1,\ (M_\star l) = 0.55$ and $T = 200$ GeV. For this plot we ignore the kinetic terms of the fields. \label{comb_pot_plot}}
	\end{figure}

Although our analysis followed a totally different route to model the black hole, i.e. the unconventional boundary condition Eq.~(\ref{BC}), we find it advisable to present an alternative analysis within the approach of Ref.~\cite{Cre2001}. The total free energy of the system in now given by
	\begin{equation}\label{comb_pot}
        V(\psi,T) = 
        \begin{cases}
        		V_{\text{GW}}(\psi) & \psi\geq0 \\
		V_{\text{BH}}(\psi,T) & \psi<0
        \end{cases}
        + \frac{\pi^4}{2}(M_*l)^3T^4\ ,
        \end{equation}
where the second term subtracts the energy of the false vacuum. Here, $V_{\text{GW}}(\psi)$ is given by Eq.~\eqref{finLag} and the black hole free energy has the form~\cite{Cre2001}
	\begin{equation}
	V_{\text{BH}}(T_h,T) = (M_\star l)^3 \pi^4 \left( \frac{3}{2}T_h^4 - 2\, T\, T_h^3 \right) \  \ .
	\end{equation}
A plot of the combined potential is seen in Figure~\ref{comb_pot_plot}. With this potential, we now have a description linking the true and false vacua, allowing a traditional bounce solution approach and removing the need for the boundary condition of Eq.~(\ref{BC}). As before, the position of the radion minimum is set to be $\mu_-=1$ TeV while that of the black hole side is simply given by $T$. Let us pause and comment on the traditional situation in the absence of brane-localised curvature. In that case both kinetic terms scale as $N^2$ and thus the contribution of each phase to the tunneling exponent strongly depends on the {\em width} of the potential on both sides at the time of tunneling. Since $\mu_-=1 \text{TeV}$ and $T_c \ll 1 \text{TeV}$, the radion motion is sufficient in analyzing the tunneling process. This simple line of reasoning fails in our model because the brane-localized curvature suppresses the dependence of the radion kinetic term on $N^2$, and thus we have to perform a full-fledged analysis of the tunneling problem. Thus the question becomes to what extent we can raise $N$ and still tunnel, while taking the black hole into account in accordance with the prescription of Ref.~\cite{Cre2001}.

We take the $T_h$ field to have a kinetic term with prefactor $(Ml)^3$ \cite{Nar2007}. The differing kinetic terms of the two fields change a number of important aspects of the bubble energy calculation which we outline. First, the critical temperature becomes		      	\begin{align}\label{Tcrit_comb}
        T_c &= \left( \frac{-2V_\text{GW}(\mu_-)}{\pi^4 \mathcal{Z}^2} \right)^{1/4},
	\end{align}
where, compared to Eq.~(\ref{Tcrit}), the dependence on $(M_\star l)^{-3/4}$ is replaced by a $\mathcal{Z}^{-1/2}$ dependence.

\begin{figure}[tbp]
	\center
  	\includegraphics[scale=0.9]{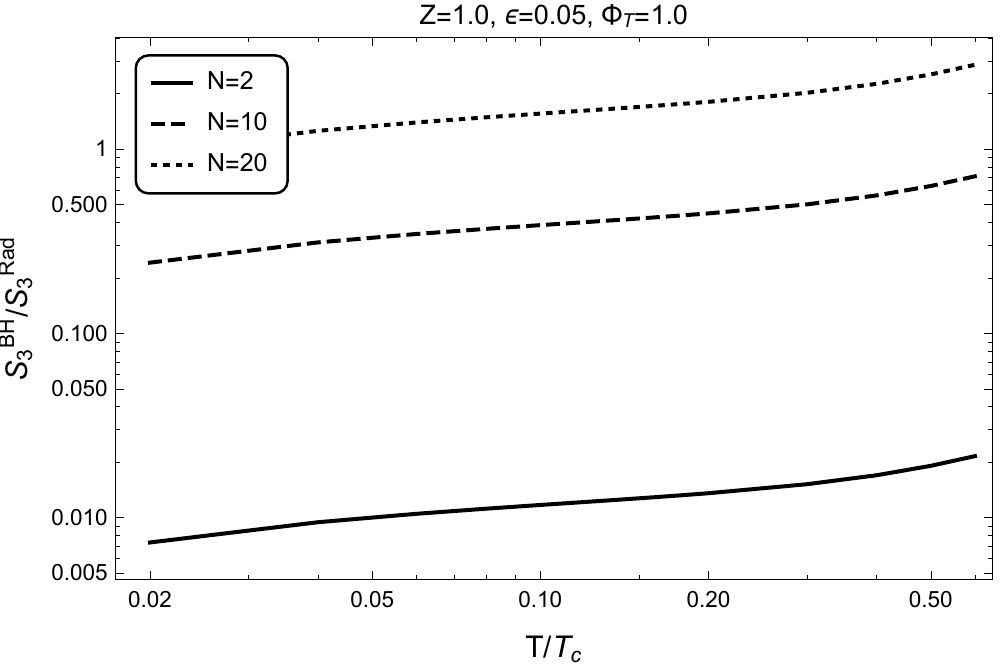}
  	\caption{A comparison of the black hole and radion contributions to the bubble energy $S_3/T$, as a function of temperature, for $N=2,10,20$ (corresponding to $(M_\star l)\sim0.4,1.4,2.2$ respectively). Any nucleation requirements are ignored in this plot for demonstrative purposes. \label{BH_Rad_ratio_plot}}
	\end{figure}

Next we have the canonically normalized $O(3)$ equation of motion that takes the form
	\begin{equation}\label{O3_eom_comb}
	\partial_r^2 \psi + \frac{2}{r}\cdot\partial_r \psi = 
	\begin{cases}
		\frac{1}{\mathcal{Z}^2}V'_{\text{GW}}(\psi) & \psi\geq0 \\
		\frac{1}{(M_\star l)^3}V'_{\text{BH}}(\psi,T) & \psi<0,
	\end{cases}
	\end{equation}
where $V'\equiv\partial V/\partial\psi$. As per usual, an overshoot-undershoot method is employed until the correct bubble solution linking the two vacua is found. The latter is then used to calculate the corresponding bubble energy. To glean the physics, we separate the tunneling exponent into radion and black hole pieces
	\begin{align}
	\frac{S_3}{T}=\frac{S_3^{\text{Rad}}}{T}+\frac{S_3^{\text{BH}}}{T}
	\end{align}
with
	\begin{align}
	\frac{S_3^{\text{Rad}}}{T} &= \frac{4\pi}{T}\cdot\mathcal{Z}^2\cdot\int_{0}^{r_{\psi=0}} dr\ r^2 \left[ \frac{1}{2}(\partial_r\psi)^2 + \frac{1}{\mathcal{Z}^2}\cdot V(\psi,T)\right]\label{S3_rad}\\
	\frac{S_3^{\text{BH}}}{T} &= \frac{4\pi}{T}\cdot(M_\star l)^3\cdot\int_{r_{\psi=0}}^{\infty} dr\ r^2 \left[ \frac{1}{2}(\partial_r\psi)^2 + \frac{1}{(M_\star l)^3}\cdot V(\psi,T)\right]\label{S3_bh}\ .
	\end{align}
Studying Eqs.~\eqref{Tcrit_comb}-\eqref{S3_bh} the dependence on some important parameters becomes clear, in particular $\mathcal{Z}$ and $(M_\star l)$ (or $N$). Just as before, decreasing $\mathcal{Z}$ reduces radion energy and enhances tunneling. Nevertheless, decreasing $\mathcal{Z}$ introduces a new feature by {\em raising} the critical temperature, which, as we discussed, increases the black hole contribution to the exponent. On the other hand $(M_\star l)$ only affects the black hole, in particular through the $(M_\star l)^3$ factor in Eq.~\eqref{S3_bh}. In Figure~\ref{BH_Rad_ratio_plot}, we show how the black hole contribution can quickly become dominant as $N$ increases from 2 to 20 (corresponding to a change in $(M_\star l)$ of just $\sim0.4 - 2.2$ ). It therefore becomes particularly important to include black hole contributions as we investigate large $N$ values.

\begin{figure}[tbp]
	\center
  	\includegraphics[width=0.49\textwidth]{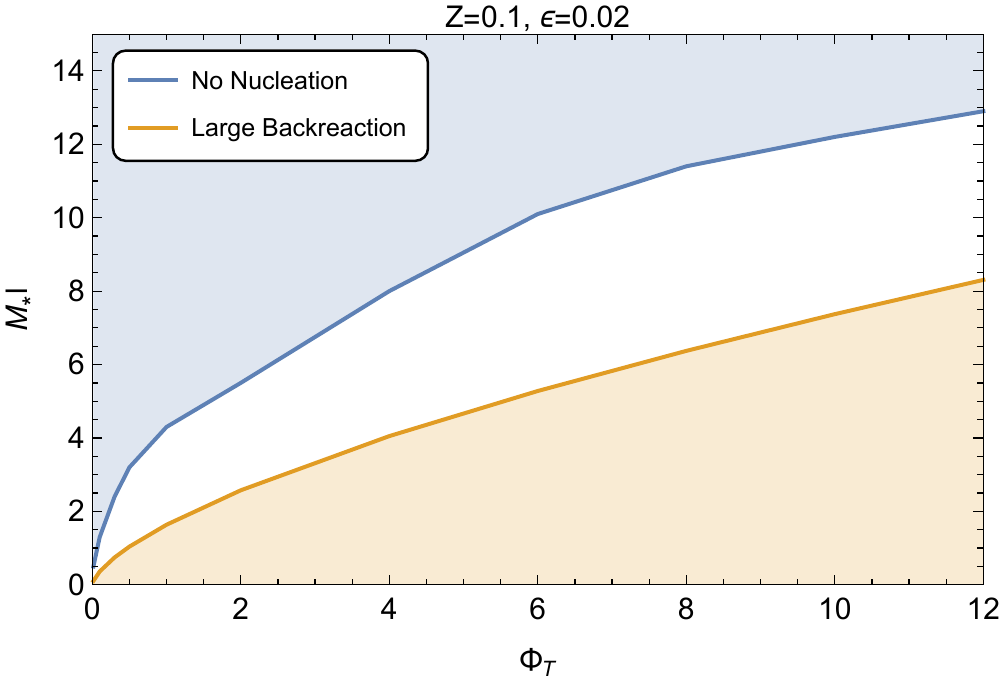} 
	\hfill
	\includegraphics[width=0.49\textwidth]{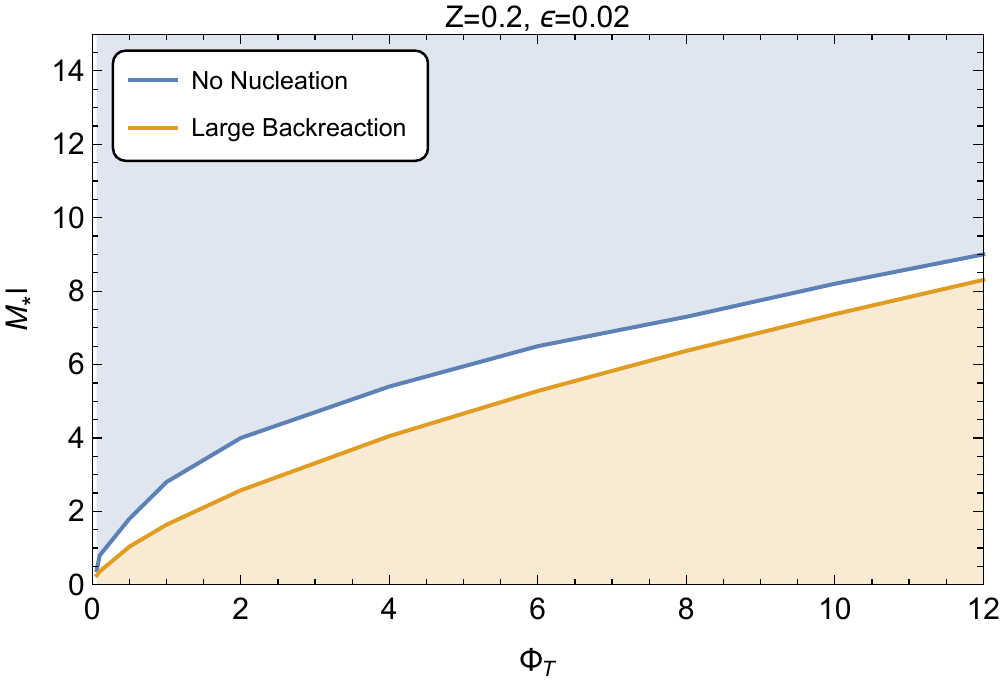}
	\vfill
	\includegraphics[width=0.49\textwidth]{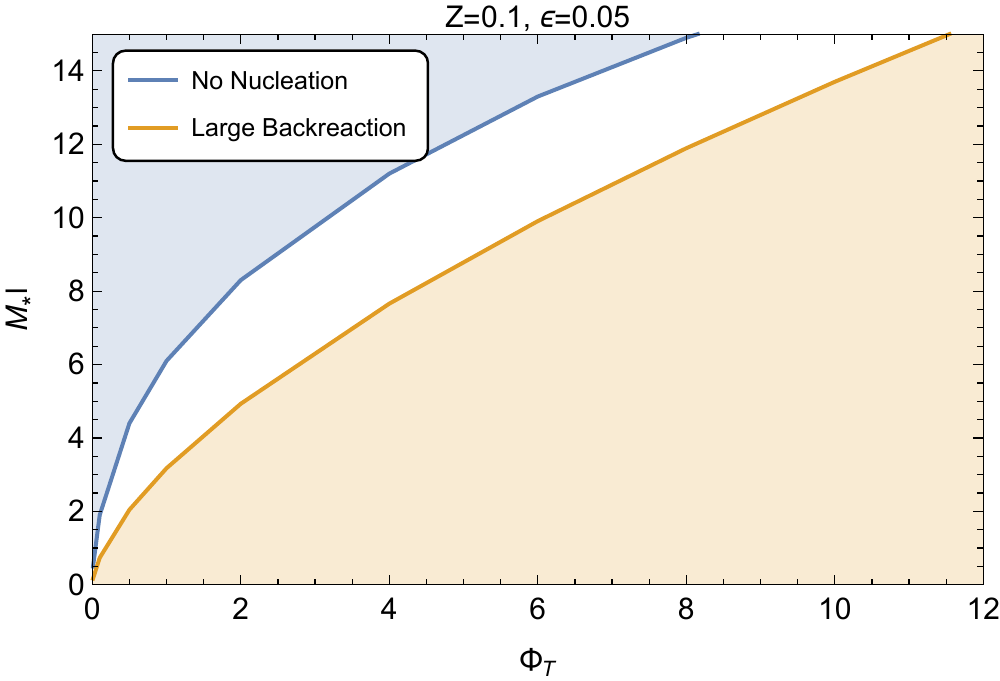}
	\hfill
	\includegraphics[width=0.49\textwidth]{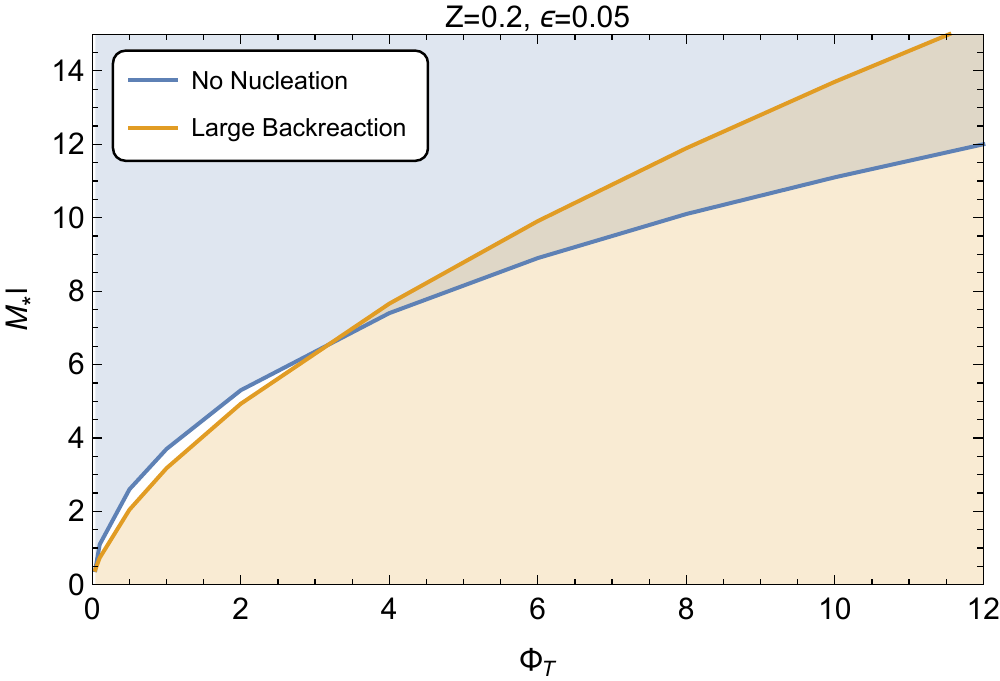}  
  	\caption{The parameter space of $(M_\star l)$ and $\Phi_T$ for varying $\epsilon$ and $\mathcal{Z}$ values for the combined potential method. The blue-shaded region represents values at which nucleation is never achieved while the orange-shaded region represents values at which back-reaction is no longer negligible. The backreaction limit is found by setting $\Phi_P=(M_\star l)^{3/2}$ in Eq.~\eqref{extrema} and solving for $(M_\star l)$. Critical temperature lines, as in Figure~\ref{parspacez1}, are no longer shown given the lack of $(M_\star l)$ dependence in Eq.~\eqref{Tcrit_comb}. \label{par_space_comb}}
	\end{figure}

To answer our basic question, we finally investigate the parameter space available in our model to achieve tunneling fully taking into account the black hole phase. Initial work made it immediately obvious that there is no available parameter space for $\mathcal{Z}\sim1$. How much then should we suppress $\mathcal{Z}$? Remarkably, Figure~\ref{par_space_comb} shows that a reduction by a factor of 10 is enough to yield a reasonable region of parameter space, where tunneling takes place at considerably large values of $(M_\star l)$. Whereas previously we found a parameter space that rapidly opened as $\Phi_T$ was increased, we now observe a bounded region available only in a range of $\Phi_T$ values.

\section{Phenomenology of brane-localized curvature}\label{sect5}

In this section, we discuss the phenomenological signatures of localized brane curvature. After the phase transition is completed, the radion classically rolls down its potential until it reaches the minimum at $\mu_-$. Fluctuations around the minimum then comprise spin-0 quanta that interact with matter on the TeV brane. We also study the KK spin-2 spectrum which is highly dependent on the new parameter $\theta_{\text{IR}}$. 
		
	\subsection{Radion phenomenology}

The first aspect one inquires about is the mass of the canonically normalized quanta, which is simply given by the second derivative of the potential at the minimum. One finds 
	\begin{align}\label{radionmass}
	\left(\frac{m_r}{\mu_-}\right)^2 &= \frac{2}{\mathcal{Z}^2}\, \epsilon (4+2\epsilon) \left(2 \epsilon \mathcal{F}^2 - 4 \mathcal{F} - \epsilon \mathcal{F} + 4 \mathcal{F}^2\right)\Phi_T^2 \nonumber \\
	&\approx\frac{16  \Phi_T^2}{\mathcal{Z}^2} \epsilon^{3/2} \ \ ,
		\end{align}
	where
	\begin{align}
	\mathcal{F} = \left[\frac{(4+\epsilon) + \sqrt{\epsilon(4+\epsilon)}}{2(2+\epsilon)}\right] \ \ .
	\end{align}

We can see that $\theta_{\text{IR}}$ has a significant effect on the mass of the radion. Without brane curvature we have that $\mathcal{Z}^2=6(M_\star l)^3$ and the upper bound on $\Phi_T^2$ is also $ (M_\star l)^3$; hence, the radion mass will be at most $\mathcal{O}(\epsilon^{3/4} \mu_-)$. On the other hand, with $\mathcal{Z}\sim\mathcal{O}(1)$ the radion mass is $\mathcal{O}(\epsilon^{3/4}(M_\star l)^{3/2}\mu_-)$, which can easily be of the order $\mu_-$ or larger depending on the precise values chosen. In fact, in most of the allowed parameter space, Figure \ref{parspacez1}, we see that the radion is order a few TeV.

	Let us now consider the couplings of the radion to a real scalar field $\phi$ on the IR brane. From Eq.~\eqref{scalarpert}, we find
	\begin{align}
	\mathcal{L}_\phi = \frac12 \left(\frac{\mu}{k}\right)^2 \left( \partial_\mu \phi\, \partial^\mu \phi - \left(\frac{\mu}	{k}\right)^2 m_0^2 \phi^2 \right) \ \ .
	\end{align}
	Now after the radion picks a vev, we canonically normalize the scalar field and define the physical mass as follows
	\begin{align}
	\phi \to \frac{\mu_-}{k} \hat{\phi}, \quad m = \frac{\mu_-}{k} m_0  \ \ .
	\end{align}
	We examine fluctuations around the radion vev as $\hat{\mu}=\langle \hat{\mu}\rangle+\hat{r}$, where $\langle\hat{\mu}\rangle\equiv \mu_-$ and $\hat{r}$ denotes the canonically normalized field with mass given in Eq.~\eqref{radionmass}. We find that the interaction between this fluctuation and the brane scalar field is given by
	\begin{align}
	\mathcal{L}_{\text{int.}} = - \frac{1}{\mathcal{Z}}\frac{\hat{r}}{\mu_-} \, T_\mu^{~\mu} \ \ ,
	\end{align}
	 where our convention for the energy-momentum tensor reads
	\begin{align}
	T_{\mu\nu} = \partial_{\mu} \hat{\phi}\, \partial_{\nu} \hat{\phi} - \frac12 \eta_{\mu\nu} \left(\partial_\lambda \hat{\phi}\, \partial^\lambda \hat{\phi}   	- m^2 \hat{\phi}^2\right) \ \ .
	\end{align}

The above derivation is quite general; the radion couples to the trace of the energy momentum tensor constructed from canonically normalized fields and physical masses.
What is important is the effect that the brane localized curvature has on the interaction strength. Here, we see that the effective interaction scale can be defined as $\Lambda_{r} \equiv \mathcal{Z}\mu_-$, whereas in the absence of brane curvature it is given by $\Lambda^{\prime}_r=\sqrt{6}(M_\star l)^{3/2}\mu_-$.
Thus, despite having large values of $(M_\star l)$, the radion still couples to matter with $\Lambda^{-1}_r\sim 1/\text{TeV}$ if we fix the brane curvature such that $\mathcal{Z}\sim\mathcal{O}(1)$.  

With the standard model residing on the TeV brane, the radion phenomenology depends only on two parameters; the mass and the effective interaction scale.
Much work has already been done, using results from LHC searches, to put experimental constraints on these two parameters \cite{Ahmed:2015uqt,Chakraborty:2017lxp,Angelescu:2017jyj}. Note that our discussion so far is particularly simple in that we do not consider a possible non-minimal coupling of the Higgs to the Ricci scalar, which should in principle be present. To estimate the constraints on $\Lambda_r$ and $m_r$ we can use the results from Ref.~\cite{Chakraborty:2017lxp}. The principle decay modes of the radion are to $W^+W^-$, $ZZ$ and $hh$, with the $ZZ\rightarrow 4l$ channel being the most constraining. Ignoring the Higgs-radion mixing, the results in \cite{Chakraborty:2017lxp} reveal that the bounds on $\Lambda_r$ range from $\sim10$ TeV for $m_r=200$ GeV to $\sim4$ TeV for $m_r=1$ TeV. Therefore with the radion mass below $1$ TeV we would expect $\Lambda_r=\mathcal{Z}\mu_-\gtrsim4$ TeV. 
The results of the previous sections do not depend strongly on the exact value of $\mu_-$; therefore, we can easily evade collider bounds by increasing the $\mu_-$ scale to $\mathcal{O}(4)$ TeV and still obtain a strong first order phase transition.
Note that varying $\mu_-$ by an $\mathcal{O}(1)$ amount will have a mild effect on the phase transition parameter space, as can be seen from the nucleation condition in Eq.~\eqref{S3T}.
A more elegant option may be to include a curvature-Higgs mixing term, in which certain values of the coupling constant can allow for lower bounds on $\Lambda_r$.

The curvature-Higgs mixing can significantly change the phenomenology of the radion; this mixing arises from a coupling of the Higgs to the Ricci scalar of the form
\begin{align}
\mathcal{L}\supset\int_{-L}^L dy \sqrt{g_L}~\delta(y-L)\xi_{hR}|H|^2R(g_L)
\end{align}
where $H$ is the Higgs doublet. For a particular value of the $\xi_{hR}$ coupling known as the conformal point, the radion couplings to the SM fields are significantly suppressed and the bounds on the effective interaction scale are reduced.
Again using the results from Ref.~\cite{Chakraborty:2017lxp} (Figure 9) we can estimate that when $\xi_{hR}$ takes the value associated with this conformal point, the experimental bound on $\Lambda_r$ is of order $1$ TeV for radion masses in the range $0.2-1$ TeV.
A dedicated study of the radion phenomenology with a larger range of radion masses and including both the brane curvature and the curvature-Higgs mixing is required in order to obtain a complete and accurate account of the bounds on the model considered in this paper; however, this is beyond the scope of our present discussion.
	
	\subsection{Spin-2 phenomenology}

\begin{figure}
	\center
  	\includegraphics[scale=0.5]{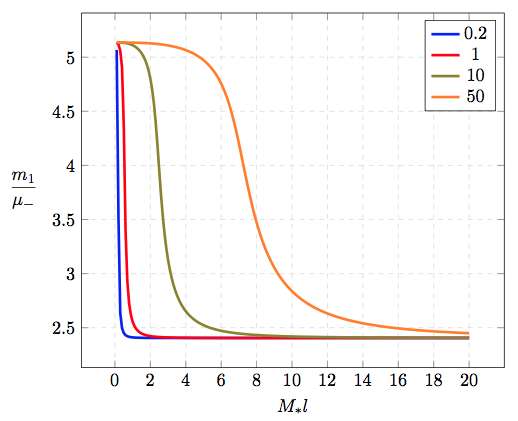}
  	\caption{Here we show how the lightest KK graviton mass varies as we change $(M_\star l)$, while constantly modifying the brane curvature term to keep $\mathcal{Z}$ at some fixed value.  We consider values of $\mathcal{Z}=0.2,1,10,50$. }
  	\label{KKgravMasses}
	\end{figure}

In this section we study the spin-2 fluctuations of the metric, which we parametrize as
\begin{align}\label{spin2pert}
ds^2 = e^{-2ky} \left(\eta_{\mu\nu}+\tfrac{1}{M_\star^{3/2}}h_{\mu\nu}\right) dx^\mu dx^\nu - dy^2, \quad h_{\mu\nu} \equiv h_{\mu\nu}(x^\mu,y) \ \ .
\end{align}
Expanding this spin-2 field in a KK decomposition, $h_{\mu\nu}=\tfrac{1}{\sqrt{L}}\sum_{n=0}^{\infty}f_n^g(y)h_{\mu\nu}^{(n)}(x)$, we find that in the bulk the 5D profile must obey \cite{Davoudiasl:1999jd}
\begin{align}\label{bulkGeom}
\partial_y^2f^g_n-4k\partial_yf^g_n+m_n^2e^{2ky}f^g_n=0 \ \ .
\end{align}
The above eigenvalue equation is obtained after a partial integration in the $y$-variable, which also generates brane localized terms $\sim f_n^g\partial_yf_n^g$. The boundary conditions are then determined by requiring that these terms are zero. However, the effects of the IR brane curvature must be accounted for here. Assuming that the KK gravitons are produced on-shell we can replace the second derivative of the fields, generated by the brane curvature, with their KK masses.
The procedure we follow is outlined in Ref.~\cite{Dav2003}.
The condition that a massless graviton exists is simply $\partial_yf_n^g=0$ on each of the branes, while for the massive modes we have
\begin{align}
\label{bulkGbcs}
\nonumber
    \partial_yf^g_n\Big\vert_0&=0  \ \ , \\
     \left(e^{-2kL} \partial_yf^g_n-\tfrac{\theta_{\text{IR}}}{2}\tfrac{m_n^2}{k}f^g_n\right)\Big\vert_L&=0 \ \ . 
\end{align}
The solution to Eq.~\eqref{bulkGeom} is
\begin{align}\label{bulkGsol1}
f^g_n=\frac{e^{2ky}}{N_n}\left(J_2\left(\tfrac{m_n}{k}e^{ky}\right)+\alpha_nY_2\left(\tfrac{m_n}{k}e^{ky}\right)\right)
\end{align}
where $\alpha_n$ and the mass spectrum $m_n$ are determined by the boundary conditions, Eq.~\eqref{bulkGbcs}, and the normalization constant $N_n$ is determined by the orthonormality condition
\begin{align}\label{Gnorm}
\frac{M_\star^3}{2k}\int_{-L}^Ldy~e^{-2ky}f_n^gf_m^g\left(k+\theta_{\text{IR}}\delta(y-L)\right)=\delta_{mn} \ \ .
\end{align}
It is useful here to consider $\theta_{\text{IR}}$ in terms of $\mathcal{Z}^2$, Eq.~\eqref{znorm}, and inspect the KK masses as a function of $(M_\star l)$ and $\mathcal{Z}$. Applying the UV boundary condition fixes
\begin{align}
\alpha_n =-\frac{J_1\left( \tfrac{m_n}{k} \right)}{Y_1\left(\tfrac{m_n}{k} \right)} \ \ ,
\end{align}
while the IR boundary condition requires
\begin{align}
J_1\left(\tfrac{m_n}{k}e^{kL}\right)-\tfrac{\theta_{\text{IR}}}{2} \tfrac{m_n}{k}e^{kL} &J_2\left(\tfrac{m_n}{k}e^{kL}\right) =\nonumber \\
& - \alpha_n  \left( Y_1\left(\tfrac{m_n}{k}e^{kL}\right)-\tfrac{\theta_{\text{IR}}}{2}  \tfrac{m_n}{k}e^{kL} Y_2\left(\tfrac{m_n}{k}e^{kL}\right) \right) \ \ .
\end{align}
Since we expect $m_n \ll k $ for the lowest-lying states and $Y_a(x)$ diverges at $x\rightarrow0$, we can approximately take $\alpha_n=0$ in the IR boundary condition.
Therefore, the masses of the spin-2 KK modes are approximated by
\begin{equation}
m_n=\text{zeros}\left[J_1\left(\tfrac{m_n}{k}e^{kL}\right)-\tfrac{\theta_{\text{IR}}}{2} \tfrac{m_n}{k}e^{kL} J_2\left(\tfrac{m_n}{k}e^{kL}\right)\right].
\end{equation}
In fact when $\theta_{\text{IR}}=1$, the spin-2 mass spectrum is exactly the same as that for spin-1 KK modes of a gauge field in the bulk of the RS model, i.e.
$m_1/\mu_-=2.40,~5.52,~8.65,~11.79$.
In Figure \ref{KKgravMasses} we show how the lightest KK graviton mass changes as we increase $(M_\star l)$ while using the brane curvature to keep $\mathcal{Z}$ at some fixed value.
The values of $\mathcal{Z}$ used here extend beyond those used in our analysis in section \ref{sect4}; however, the figure clearly shows that in the relevant parameter space the lightest KK graviton mass is significantly reduced.
We clearly see that for larger values of $(M_\star l)$ the lightest KK graviton mass approaches that of the lightest KK spin-1 mass.
This is strikingly different than in RS models without IR brane curvature, in which the lightest massive spin-2 mode is expected to be $\sim 3.8$ TeV. 

The kinetic term of the KK gravitons receives a contribution proportional to $\theta_{\text{IR}}$ thus changing their coupling to brane matter. In general, the effective KK graviton coupling is of the form
\begin{align}
S_G^{(n)}=\frac{c^{(n)}_{GX}}{M_{Pl}}\int d^4x\, h^{\mu\nu}_{(n)}T^X_{\mu\nu}=\frac{1}{(M_\star l)^{3/2}}\frac{c^{(n)}_{GX}}{k}\int d^4x\, h^{\mu\nu}_{(n)}T^X_{\mu\nu}
\end{align}
where $T^X_{\mu\nu}$ is the stress-energy tensor of some field, and the coefficient $c_{GX}^{(n)}$ depends on the 5D properties of that field.
The simplest case is matter on the TeV brane,
\begin{align}
\frac{c^{(n)}_{GX}}{M_{Pl}}=\frac{f_n^{g}(L)}{M_{Pl}\sqrt{kL}}\simeq \frac{e^{kL}}{M_{Pl}\sqrt{1+\theta_{\text{IR}}}}=\frac{1}{(M_\star l)^{3/2}\sqrt{1+\theta_{\text{IR}}}}\left(\frac{1}{\mu_-}\right) \ \ .
\end{align}
As expected, the coupling is proportional to $1/\text{TeV}$. Opposite to the radion, notice that $\theta_{\text{IR}}$ enters with a plus sign in the pre-factor of the graviton kinetic term. The effective interaction scale here can be written as $\Lambda_g=(M_\star l)^{3/2}\sqrt{1+\theta_{\text{IR}}}\, \mu_-$.
With the SM confined to the TeV brane the phenomenology of the KK gravitons is fixed by just their masses and matter couplings. Some work has been done recently to put bounds on $m_1$ and $\Lambda_g$ using the recent results from  LHC searches \cite{Dillon:2016fgw,Dillon:2016tqp,Alvarez:2016ljl}.
When both are near the TeV scale it is possible that the LHC would detect these states.

Throughout the favorable region in our parameter space, $\Lambda_g$ is required to be $\mathcal{O}(10\!-\!100)$ TeV, and thus the KK gravitons are too weakly coupled to be detected in current experiments.
From the definition of the radion interaction scale we see that, 
\begin{equation}
\frac{\Lambda_r}{\Lambda_g}=\frac{\mathcal{Z}}{(M_\star l)^{3/2}\sqrt{1+\theta_{\text{IR}}}} \ \ .
\end{equation}
The results of section \ref{sect4} tell us that we require $\mathcal{Z}\sim\mathcal{O}(1)$ and $\theta_{\text{IR}}\simeq1$ to obtain a strong first order phase transition to the RS background with large-$N$, where $N^2=4\pi^2 (M_\star l)^{3}+1$.  Therefore, at large-$N$, our model not only predicts that the radion has an $\mathcal{O}(\mu_-)$ mass and that the KK graviton mass spectrum is shifted down but also that the interaction strength of the KK gravitons is highly suppressed with respect to that of the radion.

\section{Discussion}\label{sect6}

We considered modifying the RSI theory by adding a TeV brane-localized curvature, which leaves the background solution intact, and studied the impact on the dynamics of the holographic phase transition.  The holographic interpretation is very simple: the gauge theory is de-confined at high temperature and is dual on the gravity side to the planar AdS black hole. The low temperature phase is dual to the RSI solution with two branes stabilized properly to offer a solution to the hierarchy problem. The generic features of the transition were first considered in Ref.~\cite{Cre2001}, and the physics is similar to the holographic interpretation of the Hawking-Page phase transition \cite{Wit1998,Haw1982}. 

The analysis of the phase transition is rather simplified by considering the motion of the radion in its potential that is induced by the Goldberger-Wise mechanism. Here, the contribution of the black hole phase to the tunneling rate is estimated using only the boundary condition in Eq.~\eqref{BC}, which takes into account the free energy of the black hole. It would indeed be quite valuable to try to find the full gravitational and bulk scalar instanton, or at least have a better estimate of the effect of the high temperature phase on the tunneling rate.

The brane-localized curvature contributes to the kinetic energy of the radion in a fashion that makes it possible to desensitize the dependence of the tunneling rate on the fundamental combination $(M_\star l)$. Now that $(M_\star l)$ (or $N$) can be made large, the tunneling rate is further enhanced by propping up $\Phi_T$, the Goldberger-Wise TeV-vev, consistent with back-reaction constraints. Contrary to previous studies, over a wide range of parameters we find $T_n/T_c \sim \mathcal{O}(0.1-1)$.

We end by commenting on the possibility of tunneling through $O(4)$-symmetric bubbles. Although we did not perform a detailed study, it is likely that $O(4)$ bubbles dominate over the $O(3)$ ones. The physics is gleaned by inspecting the $O(4)$ bounce action
\begin{align}\label{o4}
S_4 = E_4 + \frac{\pi^6}{4} (M_\star l)^3 (r_c T)^4 \ \ .
\end{align}
In fact, from our presented work on the $O(3)$ case we know that the temperature roughly scales as $(M_\star l)^{-3/4}$. This implies that the second term in Eq.~\eqref{o4} is uniform with respect to $(M_\star l)$. For fixed values of $(M_\star l)$ and $\Phi_T$, we then expect $S_4 < S_3/T$. This indeed does not invalidate any of our conclusions regarding the completion of the phase transition as the tunneling would be quicker with $O(4)$. An in-depth study of this possibility is left to future work.

\section*{Acknowledgements}
We would like to thank Thomas Konstandin for valuable discussions. The work of B.D. is supported by EPSRC Grant No. EP/P005217/1. The work of B.K.E. and S.J.H. is supported by the Science Technology and Facilities Council under Grant No. ST/L000504/1. J.P.M. is supported by a Science and Technology Facilities Council studentship.



\end{document}